\documentclass{elsart}
\usepackage{graphicx}
\usepackage{amssymb}
\renewcommand{\thefootnote}{\fnsymbol{footnote}}
\def \lsim {\,{\scriptscriptstyle{\stackrel{<}{\sim}}}\,}

\newcommand{\bes}{\begin{eqnarray}}
\newcommand{\ees}{\end{eqnarray}}
\renewcommand{\thesection}{\arabic{section}.}
\renewcommand{\thesubsection}{\arabic{section}.\arabic{subsection}.}

\begin{document}

\thispagestyle{empty}
\begin{frontmatter}
\title{
Precise comparison of theory and new experiment for the Casimir force
leads to stronger constraints
on thermal quantum effects and long-range interactions
}
\author[a]{R.~S.~Decca,}
\author[b]{D.~L\'{o}pez,}
\author[c]{E.~Fischbach,}
\author[d]{G.~L.~Klimchitskaya,}
\author[e,c]{D.~E.~Krause,}
\author[f]{V.~M.~Mostepanenko\corauthref{1}
}

\address[a]{Department of Physics, Indiana
University-Purdue
University Indianapolis, Indianapolis, Indiana 46202, USA }
\address[b]{
Bell Laboratories, Lucent Technologies, Murray Hill, New
Jersey 07974, USA}
\address[c]{Department of Physics, Purdue University, West Lafayette, Indiana
47907, USA,}
\address[d]{North-West Technical University, Millionnaya St. 5, St.Petersburg,
191065, Russia}
\address[e]{Physics Department, Wabash College, Crawfordsville, Indiana 47933,
USA }
\address[f]{Noncommercial Partnership ``Scientific Instruments'',
Tverskaya St. 11, Moscow, 103905, Russia
}
\corauth[1]{Corresponding author. \\E-mail address:
{Vladimir.Mostepanenko@itp.uni-leipzig.de}}

\begin{abstract}
We report an improved dynamic determination of the Casimir
pressure $P^{\rm expt}$ between two plane plates obtained using
a micromachined torsional oscillator. The main improvements
in the current experiment
are a significant suppression of the surface roughness
of the Au layers deposited on the interacting surfaces, and
a decrease by a factor of 1.7 (down to 0.6\,nm)
in the experimental error in the measurement of the absolute
separation. A metrological analysis of all data for $P^{\rm expt}$ from
15 sets of measurements permitted us to determine both the
random and systematic errors, and to find the total experimental error
in $P^{\rm expt}$ as a function of separation
at the 95\% confidence level. In contrast to all previous experiments
on the Casimir effect, where a small relative error was achieved only
at the shortest separation, our smallest experimental error
($\sim 0.5$\%) is achieved over a wide separation range.
The theoretical Casimir pressures $P^{\rm theor}$ in the
experimental configuration were calculated by the use of
four theoretical approaches  suggested in the literature
based on the Lifshitz formula at nonzero temperature.
All corrections to the Casimir force due to grain structure of the
overlying metal layers (including the variation of optical data and
patch potentials), surface roughness (including nonmultiplicative and
diffraction-type effects), and nonlocal effects, were
calculated or estimated. The maximum value of the roughness
correction, achieved at the shortest separation of 160\,nm, is equal
to only 0.65\% of the Casimir pressure.
All theoretical errors, including those
introduced by the proximity force theorem, finite size of the
plate area, and uncertainties in the experimental separations, were
analyzed and metrologically combined to obtain the total
theoretical error at the 95\% confidence level. Finally, the confidence
interval for $\left( P^{\rm theor}-P^{\rm expt}\right)$ was obtained
as a function of separation.
Our measurements are found to be consistent with two theoretical
approaches utilizing the plasma model and the surface impedance
over the entire measurement
region from 160\,nm to 750\,nm. Two other approaches to the
thermal Casimir force, utilizing the Drude model or a special
prescription for the determination of the zero-frequency contribution
to the Lifshitz formula, are
excluded on the basis of our measurements at the 99\% and 95\%
confidence levels, respectively. Finally, constraints on Yukawa-type
hypothetical interactions are strengthened by up to a factor of 20
in a wide interaction range.
\end{abstract}
\end{frontmatter}

\section{Introduction}

Recent advances in the experimental and theoretical investigation
of the Casimir effect  suggest that this field of research is
important for both fundamental and applied physics.
The Casimir force, which acts between two uncharged
plates in vacuum, is a purely quantum effect caused by
the alteration of the zero-point oscillations
of the electromagnetic field due to boundaries \cite{1}.
The modern stage in Casimir research originated in a number
of measurements of the Casimir force between metals (see
 \cite{2,3,4,5,6,7,8,9,10,11} and review \cite{12}).
Additional motivation for studying the Casimir force comes from
promising applications of these results in both fundamental
physics (for constraining hypothetical long-range interactions
\cite{11,13,14,15,16,17}), and in nanotechnology \cite{18,19}.
Recently the Casimir-Polder force has attracted attention
\cite{19a,19b} in connection with experiments on quantum reflection
and Bose-Einstein condensation.

As often happens, experimental progress placed more stringent
demands upon theoretical calculations of the Casimir force.
At present, to achieve an agreement between theory and
experiment at the level of 1\% of the measured force, special care
must be taken to account for all relevant factors, such as
the finite conductivity of the metallic boundaries, the effect
of thermal corrections, surface roughness, grain structure of the
covering layers, and patch potentials. These
are reviewed in  \cite{12}, and more recently in
 \cite{20} which is devoted to
the application of these questions to Casimir force measurements
between gold surfaces by means of an atomic force microscope
\cite{6}.

The basic theoretical formalism for the calculation of the van der
Waals and Casimir forces between real materials is given by
Lifshitz theory \cite{21,22,23}. Beginning in 2000, several
research groups applied this theory to the case of real metals
at nonzero temperature and four different approaches were proposed.
In the first approach \cite{25} the zero-frequency contribution
to the Lifshitz formula was determined by the use of the Drude
dielectric function. The second approach \cite{27}
modified the zero-frequency term of the Lifshitz formula
to make it the same for ideal and real metals.
In the third approach \cite{29,30} the Lifshitz formula was combined
with the free electron plasma model neglecting relaxation.
Finally, in  \cite{24} a fourth approach was proposed
in which the Lifshitz formula at
nonzero temperature is rederived using the Leontovich surface
impedance boundary conditions rather than the bulk dielectric permittivity.
Within this last approach \cite{24,28}, the zero-frequency term of the
Lifshitz formula is determined by the form of the Leontovich impedance
valid near the characteristic frequency. Detailed discussions of
the advantages and disadvantages of each of
proposed approaches can be found in
 \cite{31,26,54,55,38a,38b,26b,41b,41a,Lam1,Lam2}.
These discussions are part of a broader context  connected with the
fundamental nature of the zero-point oscillations and their relation to
basic physical principles such as the laws of thermodynamics.
In particular, it has been shown \cite{28,26} that the first and second
theoretical approaches to the description of the thermal Casimir
force between real metals lead to a violation of the third law
of thermodynamics (the Nernst heat theorem) in the case of perfect
crystals. The attempt made in \cite{38b,26b} to avoid this 
conclusion in the
framework of the first approach required the introduction of defects and
impurities to the crystal, which cannot be considered as satisfactory 
since the perfect crystal
  forms the basis of  the theory of electron-phonon interactions. It 
follows that any formalism
  applied to perfect crystals must satisfy the third law of 
thermodynamics, which the
  formulations of the first approach do not.

We consider now a comparison of the different theoretical approaches
with experiment.
It should be noted that the third and fourth approaches
predict small thermal corrections to the Casimir force at short
separations of order 100\,nm, in qualitative agreement with the
case of an ideal metal, whereas the first and second approaches
lead to thermal corrections which are many times greater \cite{31}.
However, taking into
account the small magnitude of the Casimir force, reliably
distinguishing among the different predicted thermal corrections
presents a serious challenge to experimental physics. In this
regard it is important that the first approach (see  \cite{25}
and also  \cite{54,55,38a,38b}) disagrees significantly
\cite{Lam1,Lam2} with the most famous modern measurement of the
Casimir force between Au surfaces by means of a torsion pendulum
\cite{2}. A net deviation from zero temperature Casimir force is
predicted by the first approach to be about 25\% at a separation
$a=1000\,$nm. It was not, however, observed \cite{2,Lam1,Lam2}
despite the fact that the experimental uncertainty was less
than 10\%.

The most precise and accurate measurement of the Casimir force to
date was made between two dissimilar metals (Cu and Au) by means of a
microelectromechanical torsional oscillator \cite{10,11}. In that
experiment the Casimir force between a plate and a sphere was
measured statically, and the effective Casimir pressure between two
parallel plates was determined dynamically, with an absolute error of
$\approx 0.6\,$mPa at 95\% confidence. The estimated theoretical
uncertainty in  \cite{11} was approximately 1\% of the Casimir
pressure over the entire measurement range. This was, however, a
purely theoretical error which did not take into account errors in
the theoretical pressure induced by uncertainties in the experimental
determination of the sphere-plate
separations. (Recall that a micromachined oscillator
was first used in  \cite{18} to demonstrate a
micromechanical device driven entirely by the Casimir force.) The
experimental data in  \cite{11} were not precise enough
to observe the traditional
thermal corrections of the third and fourth approaches.
However, the comparison of these data with theory demonstrates
that the alternative thermal corrections to
the Casimir pressure, as predicted in the first \cite{25} and
second \cite{27} approaches, are excluded by
experiment within a wide separation region.
The good agreement between theory
(based on the surface impedance approach) and experiment was used
in  \cite{11} to obtain stronger constraints on hypothetical
long-range interactions predicted by many extensions to the
Standard Model of fundamental interactions.

In this paper we present the results of an improved
experimental determination of the Casimir pressure
reported in  \cite{11}, and of a refined comparison
of these new results with all four theoretical approaches.
The following modifications in the experimental setup
allowed us to obtain results exceeding those of  \cite{11} in both
reliability and conclusiveness. The main improvement is associated
with a considerable decrease of the surface roughness on both
the plate and the sphere. In  \cite{11} the highest roughness
peaks were as much as $\approx 100\,$nm
in  height on both bodies. In the
improved experiment the maximum heights of the roughness peaks are
11.06\,nm on the sphere and 20.63\,nm on the plate.
An improvement in detection sensitivity, together with a reduction of
the coupling between the micromachined oscillator and the environment,
yielded measurements at smaller separations
between the test bodies (160\,nm instead of 260\,nm). Another
important improvement was the reduction of the error in determining the
absolute separation between the sphere and the top of the
metal-coated plate from
1\,nm in  \cite{11} to 0.6\,nm. A special experimental
study was performed to
ensure the linearity of the microelectromechanical torsional
oscillator. In addition Au was used as a covering layer on both the
plate and the sphere (instead of Cu and Au, respectively, as in
 \cite{11}), which permits a more reliable theoretical
interpretation of the experimental results.

The refined theory includes a complete analysis of the finite
conductivity, surface roughness, and thermal corrections to the
Casimir pressure. In contrast to  \cite{11}, the finite
conductivity corrections are analyzed by taking into account the grain
structure of the covering layers, possible sample-to-sample
variations of the tabulated optical  data, and the
contribution of nonlocal effects. The roughness analysis incorporates
non-multiplicative, diffraction-type and correlation effects.
It is shown that in the present experiment the maximum value
of the roughness correction, which is achieved at a separation
of 160\,nm, is only 0.65\% of the Casimir pressure.
Thermal Casimir pressures are computed in
the framework of all four theoretical approaches.

A special metrological analysis has been performed separately for the
experimental data and for the results of the theoretical
computations. The random, systematic, and
total experimental errors are found
as functions of separation, taking into account the
distribution laws for each error at the 95\% confidence level.
A careful comparison between experiment and theory has been
carried out  taking into
account all experimental and theoretical errors.
This comparison is based on rigorous metrological criteria
which do not use the concept of the root-mean-square deviation
between theory and experiment over the entire separation range,
a procedure that had been
criticized in literature \cite{7}.
As a result, the alternative thermal corrections to the Casimir
pressure, as proposed in the first \cite{25} and second \cite{27}
approaches, are excluded by experiment, whereas
the third \cite{29,30} and fourth \cite{24,28} approaches are
shown to be consistent with experiment. These results
confirm those obtained earlier in  \cite{11}.
In this
paper we also improve the constraints on hypothetical long-range
interactions obtained in  \cite{11}.
By combining the results of this paper and  \cite{11},
 the previously known constraints are strengthened by
a factor of up to 20 in a wide interaction region around 0.1\,$\mu$m.

The present paper is organized as follows. In Section 2 the experimental
configuration and sample characterization are briefly described.
Section 3 contains the calibration procedure and the experimental
results for the Casimir pressure. In Section 4 the metrological analysis
of these results is performed and the experimental errors are found at
95\% confidence. In Section 5 we calculate the Casimir pressure between
Au layers in the framework of the four approaches
described above taking into account different kinds of corrections.
Section 6 is devoted to a metrological analysis of the theoretical
uncertainties. In Section 7 all four theoretical approaches to the thermal
Casimir force are  compared with the experimental data.
The level of agreement between experiment and theory is used
to obtain stronger constraints on hypothetical long-range interactions
in Section 8. We conclude in Section 9 with a discussion
of our results.

\section{Improved experimental setup and sample characterization}

\subsection{Details of the improved experiment}
The general scheme of the experimental setup used to determine the
Casimir interaction between two bodies by means of a
micromachined oscillator was described previously in
 \cite{10,11}.  One of the new features of the present
experiment is the fact that here the Casimir attraction is
measured between two layers of Au, specifically,
a Au-coated sapphire sphere and a Au-coated polysilicon
microelectromechanical torsional oscillator (MTO). To improve the
force sensitivity, the vertical separation between the sphere and
the plate was changed harmonically with time, leading to a
measurement of the $z$-derivative of the Casimir force. As
described in  \cite{11}, this is equivalent to measuring the
Casimir force per unit area, or the Casimir pressure, for a
configuration of two parallel plates (see below).
The MTO presents a low coupling with the environment, since it is
less sensitive to modes that involve a displacement of the center
of mass. The miniaturization also allows us to achieve a large
quality factor $Q\sim 8000$.

The experimental arrangement is shown schematically in
Fig.~\ref{fig1}. The MTO is made of a 3.5\,$\mu$m thick,
500\,$\times$ 500\,$\mu$m$^2$ heavily doped polysilicon plate
suspended at two opposite points by serpentine springs. Two
independently contacted polysilicon electrodes located under the
plate are used to induce an oscillation on the plate at the
required resonant frequency of the MTO. A 80 $\mu$m wide ribbon at
the edge of the plate was coated with 10\,nm of Pt followed by
150\,nm of Au. The Au layer constitutes one of the metals used in
the measurement of the Casimir interaction. The separation between
the two Au layers is given by
\begin{equation}
z = z_{meas} - D -b\theta.
\label{e0}
\end{equation}
\noindent
 In this
expression $z_{meas}$ is the separation between the end of the
cleaved fiber and the platform, as interferometrically measured
with the absolute error $\Delta z_{meas}=0.2$\,nm. $D=D_1+D_2$,
where $D_2$ represents
the separation between the platform
and the top of the Au-coated plate
when the interaction between the sphere and the plate is
negligible (i.e., $\theta=0$), and $D_1$ is the distance between
the end of the bottom of the cleaved fiber and the
external surface of the
Au film on the Au-coated sphere (see Fig.~1).
The lever arm $b = (210 \pm 3)\,\mu$m is determined
optically. The value of $\theta$ is determined by measuring the
difference in capacitance between the plate and the right and left
electrodes $C_{r,l} = C_{right} - C_{left}$.
In all reported cases $\theta\leq 10^{-5}\,$rad.
A field effect
transistor (FET), the first stage in the amplifier circuit, was
placed as close as possible to the oscillator to minimize the
effect of parasitic capacitances. The smallest angular deviation
that can be detected in this configuration is given by

\begin{equation}
 \Delta \theta \approx \frac{\Delta C_{r,l}}{C_o}\frac{2
d_g}{w'}\approx \frac{\Delta V}{V_i}\frac{C_T}{C_o}\frac{2
d_g}{w'} \sim 10^{-9} \sqrt{\nu}\frac{\rm rad}{\sqrt{\rm Hz}}, \label{eq1}
\end{equation}

\noindent where $\Delta C_{r,l}$ is the minimum detectable change
in $C_{r,l}$, and $C_o \approx {\epsilon_o w w'}/{d_g}$ is the
capacitance between the oscillator and each electrode when no
forces are present ($\epsilon_o$ is the permittivity of free space).
The dimensions of the electrodes are $w = 250
\mu$m, $w' = 190 \mu$m, and $d_g =2 \mu$m is the separation
between the plate and the electrodes. $C_T \approx$ 20\,pF $\gg
C_o$ is the parasitic capacitance of the measurement circuit
(determined by the capacitance of the FET transistor plus
parasitic capacitances to ground),
$\Delta V \sim 10\sqrt{\nu}\,$nV/Hz$^{1/2}$
is the input noise of the amplifier, $\nu$ is the bandwidth
of the measurement, and
$V_i\>(i=1,2) \sim$ 1 V is the DC potential used to correct for initial
asymmetries in the circuit and to linearize its response. In
obtaining Eq.~(\ref{eq1}) parallel plate approximations, and the
fact that the maximum angular deviation $\theta_{max} \ll 1$, have
been used. In the relevant $(100-5000)\,$Hz range, actual
measurements
of $\Delta C_{r,l}$ coincide within 15\,\% with those predicted by
Eq.~(\ref{eq1}) .

A sphere with a nominal radius $R = 150\,\mu$m was coated with a
10\,nm layer of Ti followed by a 200\,nm layer of Au.
For both the plate of the
MTO and the sphere, vacuum was not interrupted in between
depositions. Deposition induced asymmetries in the sphere were
found to be smaller than 10\,nm, the resolution of the scanning
electron microscope (SEM) used to characterize them. The entire
setup (MTO and fiber-sphere) was rigidly mounted into a can which
was hung inside a vacuum chamber by means of soft springs. A
built-in magnetic damping vibration isolation helped in reducing the
coupling with the main resonance of the building, occurring at
7\,Hz. Finally, the vacuum chamber was, in turn, mounted onto an
air table. This combination of vibration isolation systems yielded
peak-to-peak vibrations with $\Delta z^{pp} < 0.02$\,nm (the
detection limit in our accelerometer) for frequencies above
100\,Hz. To reduce the damping in the MTO, the vacuum chamber was
initially evacuated to a pressure less
than $10^{-5}$\,torr. In order to reduce external vibrations,
the chamber was then closed, the pumping station
removed, and the vacuum lines disconnected. The pressure was kept
low during the experiment by means of a chemical pump, made of a
cold ($T = 77$K) activated carbon trap located inside the vacuum
chamber. A pressure $\lsim 10^{-5}$\,torr was maintained with this
setup for the length of each experimental run. When necessary,
the vacuum chamber was re-evacuated between runs.

\subsection{Sample characterization}
Before performing the Casimir interaction measurements, the sample
was characterized using an atomic force microscope (AFM). This
serves the double purpose of determining the roughness of the
sample and also, by repeating the AFM study {\it after} the
Casimir measurements, to check that the sample was not modified
during the measurement. The AFM images were obtained for both the
plate and the sphere. The results are shown in Fig.~\ref{AFM}. It
is not possible for us to ensure that the region shown in
Fig.~\ref{AFM}a and Fig.~\ref{AFM}b (before and after the
measurements, respectively) is exactly the one giving
 the dominant contribution to the
interaction with the sphere. We note, however, that we
imaged with the AFM a region of $(100 \times 100)\,\mu\mbox{m}^2$,
located at the center of the Au-coated region of the plate, and
Fig.~\ref{AFM}a represents just a fraction of this region.
The remainder
of the investigated region, however, shows a very similar
behavior, with identical topography except for two dust particles
observed. For the sphere it is easier to find the interacting area
since the fiber itself provides a preferential direction.
Figs.~\ref{AFM}c and \ref{AFM}d
show AFM images of the bottom of the sphere,
obtained before and after the experiment. As in the case of the
plate, there are no discernible modifications in the topography.
It is worth mentioning that a spherical surface of radius $R$ was
first subtracted from the data shown in Fig.~\ref{AFM}c and
Fig.~\ref{AFM}d. For all cases shown in Fig.~\ref{AFM} an extra
planarization was performed to account for the small
non-linearities of the {\it z}-axis of the AFM's translational
stage. The extra planarization for both the plate and sphere had
$|\Delta z| < 0.5\,$nm. In Fig.~\ref{f3}, typical cross-sections
of the images of Figs.~\ref{AFM}b and \ref{AFM}d
at $y=\mbox{const}$ are shown for
the plate (a) and the sphere (b).
Results obtained either before the Casimir measurements or at
$x=\mbox{const}$ are quite similar. Fig.~\ref{f3} permits us to
estimate the correlation length of surface roughness (see
Section 5.4).

These AFM images were used to characterize the
roughness of the sample, which has to be taken into account when
comparing the measured Casimir interaction with the theoretical
predictions. As was done in  \cite{11}, the roughness is
represented by the fraction $v_i$ of the sample with height
$h_{i}$. Data resulting from the AFM images of Figs.~\ref{AFM}b
and \ref{AFM}d are presented in Figs.~4a and 4b, respectively. The
heights $h_i$ are plotted along the vertical axis as a function of
the fraction $w_i$ of the total surface area having height $h <
h_{i}$. The width of each horizontal step is equal to the fraction
of the total area $v_i$ with heights $h_i\leq h < h_{i+1}$ ($1\leq
i\leq K$, where $K=105$ for the plate and $K=112$ for the sphere;
in both cases $h_1=0$). Evidently $w_i=v_1+v_2+\ldots+v_i$ and
$w_K=1$. The data of Figs.~4a and 4b will be used in Section 5.4 for
the calculation of the roughness corrections to the Casimir
pressure. As was noted in the Introduction, the surface roughness
was considerably decreased compared with  \cite{11}. (The
highest peaks here, $h_{112}=11.06\,$nm on the sphere and
$h_{105}=20.65\,$nm on the plate, are much lower than the highest peak
$h_{59}=98.5\,$nm in  \cite{11}.)

\section{Experimental results for the Casimir pressure}

\subsection{System calibration}

The calibration of the system was performed in an analogous manner
to that reported in  \cite{11}. The electrostatic force $F_{el}(z)$
between the plate and the sphere was measured as a function of the
separation $z$ for $z > 3\,\mu$m, where the Casimir force is smaller
than 0.1\% of the total force. This measurement was repeated for
different potential differences between the plate and
the sphere. For a given potential difference, $F_{el}(z)$ between a sphere
and an infinite plane is given by \cite{31a}

\begin{equation}
F_{el} = 2\pi\epsilon_0(V_{Au} - V_0)^2 \sum_{n=1}^{\infty}
\frac{\coth (u) - n \coth (nu)}{\sinh (nu)} \label{eqe1}.
\end{equation}

\noindent Here
$V_{Au}$ is the voltage applied to the sphere, $V_0$ is the
residual potential difference between the metallic layers when
they are both grounded, and $\cosh u = [1+z/R]$. Fig.~\ref{deltaV}
shows the dependence of $\theta$ (and hence the electrostatic
force) on the applied voltage $V_{Au}$. Adjusting $V_{Au}$ in
Eq.~(\ref{eqe1}) we found that $F_{el} = 0$\,N within the
experimental uncertainty for $V_0 = (17.5 \pm 0.1)$\,mV. $V_0$
represents the difference in work functions between the Au layer
on the plate and the sphere, along the path that closes the
electric circuit between them. This value was observed to be
constant for $z$ in the (0.15--5)\,$\mu$m range, and it did not vary
when measured over different locations in the Au layer on the
plate. Note that $F_{el}$ is measured between the zero roughness levels
on a plate and a sphere relative to which the mean values of
roughness profiles are zero (see also Section 5.4). As a result, the
absolute separations from Eq.~(\ref{e0}) are also determined between
the zero roughness levels. For this reason, there is no systematic error
due to roughness in the measurements of separations in addition to
uncertainties discussed in Section 4.1 below.

A set of 120 curves of $F_{el} (z)$ was then used to  fit
for the parameter $D=D_1+D_2$ (the
error on $D$ was originally estimated to be no smaller than 50 nm).
To check the stability of the system, a
subset of these curves (typically 30 of them) was repeated after
each set  of the separation dependence measurements,
$P(z)$, of the Casimir interaction.
Three curves for different $\Delta V$ are shown in
Fig.~\ref{Fel}. The local $R$ in the configuration when the sphere
and the plane are interacting, and the proportionality factor $k$
between the measured $\Delta C$ and $\theta$, are also obtained
through the measured $F_{el} (z)$.
Finally, the following values of all three parameters were determined:
$D=(9349.7\pm 0.5)\,$nm, $R=(148.7\pm 0.2)\,\mu$m,
and $k=(50455\pm 7)\,$N/F.
The determination of $D$ and
the measurement of $\theta$ then yield $z$ when $z_{meas}$ is
measured interferometrically.

\subsection{Casimir pressure measurements}

As in  \cite{10,11}, the most sensitive
measurement of the Casimir interaction arises from measuring
the change of the angular resonant frequency $\omega_r$ of the
oscillator in the presence of vacuum fluctuations. In this
situation, $\omega_r$ is given by \cite{10,11,18}

\begin{equation}
\omega_r^2 = \omega_0^2 \left [
1-\frac{b^2}{I\omega_0^2}\frac{\partial F}{\partial z}\right ],
\label{approx}
\end{equation}

\noindent where $\omega_o=2\pi\times 702.92\,$Hz is the natural
angular resonant frequency of the MTO, $F$ is the Casimir force
between a plate and a sphere, $I$ is the moment of inertia of the
MTO, and ${b^2}/{I} = (1.2579 \pm 0.0006)\,\mu\mbox{g}^{-1}$ was
obtained from the electrostatic interaction $F_{el}$. Using the
proximity force theorem, we arrive at the Casimir pressure between
the two plates
\begin{equation}
P(z)= -\frac{1}{2 \pi R}\frac{\partial F(z)}{\partial z}.
\label{e3a}
\end{equation}

The resonant frequency of the MTO at a separation ${z}$
[and, hence, $P(z)$] was measured by changing the
separation between the sphere and the plate as

\begin{equation}
 \tilde{z} (t) = {z} + A_z \cos(\omega_r t),\label{eq2}
\end{equation}

\noindent where $A_z/z \ll 1$. For appropriately small values of
$A_z$ Eq.~(\ref{approx}) is recovered. In reality, however, this
is only an approximation, and the actual equation of motion for
the MTO becomes non-linear \cite{18}. For the determination of
$\omega_r$ to be experimentally relevant, the value of $A_z$ has
to be sufficiently small such that the effect of non-linearities
is smaller than $\Delta \omega_r$, the error in the measurement of
$\omega_r$. On the other hand, $A_z$ should be as large as
possible to minimize $\Delta \omega_r$. The empirical approach we
used in this paper is the following: At a separation ${z}$ we
determined $\tilde{\omega}_r$ by measuring the spectral response of
$C_{r,l}$ under the influence of thermodynamic noise. These
measurements were performed with an integration time $\tau = 500\,$s.
Subsequently a motion described by Eq.~(\ref{eq2}) was induced
with $\omega = \omega_r$ fixed by means of a phase-lock-loop
circuit. This was done while increasing the value of $A_z$ until a
decrease in $\Delta \omega_r$ was observed, while
at the same time holding $\omega_r$ fixed at
 the value obtained by measuring the spectral
response of the MTO. These results, shown in Fig.~\ref{freq},
allowed us to define the maximum value of $A_z$ to be used
at different ${z}$. The minimum value used was $A_z=1.2\,$nm at
$z=160\,$nm.

The absolute error of $\omega_r$, $\Delta \omega_r$, was found to
be $2 \pi\times 6\,$mHz
for an integration time of 10\,s. This is approximately
a factor of 1.7 smaller
than in our previous work \cite{11}, the main
reasons being a better decoupling of the apparatus from the
environmental noise, and an improvement in the amplifier to
measure $C_{r,l}$.

The Casimir pressure $P(z)$
was measured 15 times
over different positions on the sample within a separation region
$z=(160-750)\,$nm, and one of the data sets is
shown in Fig.~\ref{data}. Each point in the figure was obtained
with an integration time of 10 s.

\section{Metrological analysis of experimental errors}

\subsection{Random errors}
We start from the determination of the absolute error $\Delta z$
in the separation $z$ measured between the zero roughness
levels of Au films on
the plate and on the sphere. According to Eq.~(\ref{e0}),
$\Delta z$ is
determined by the absolute errors $\Delta z_{meas}=0.2\,$nm and
$\Delta D=0.5\,$nm.
(We note from Eq.~(\ref{eq1}) that with
$\Delta\theta\approx 10^{-9}\sqrt{\nu}\,$rad/Hz$^{1/2}$ and
$\theta\leq 10^{-5}\,$rad \cite{11}, the third term in the
right-hand side of Eq.~(\ref{e0}) makes a negligible contribution to
$\Delta z$.) Both  $z_{meas}$ and $D$ are assumed to be distributed
uniformly within the limits of their respective absolute errors
(note that the use of other distributions would decrease $\Delta z$).
The absolute error of the quantity obtained by the composition of the
uniform distributions is given by \cite{32}
\begin{equation}
\Delta z=\mbox{min}\left[(\Delta z_{meas}+\Delta D),\,
k_{\beta}\sqrt{(\Delta z_{meas})^2+(\Delta D)^2}\right],
\label{e5}
\end{equation}
\noindent
where the correction factor $k_{\beta}$ depends on the chosen confidence
level and the number of composed quantities (two in our case).
Here and below we will use the confidence level
$\beta=95$\% which results in $k_{\beta}=1.1$
\cite{32}. Substituting this into Eq.~(\ref{e5}) we conclude that
the absolute separations are measured with an error $\Delta z=0.6\,$nm.

As mentioned in Section 3.2, the Casimir pressure was measured
15 times over the 160\,nm to 750\,nm separation region
(with  288 to 293 points in each set of
measurements). It is not trivial, however, to take advantage of
the fifteen-fold repetition of the measurement in order to
decrease the random error and to narrow the confidence interval.
The difficulty is that in the measurement procedure employed the
separation step between any two neighboring points was not uniform,
even in one set of measurements, and
additionally was quite different in all sets of the data.
This leads to an absence of even a few points (let alone fifteen) taken
at  the same separation which could thus be used for averaging in
a statistical analysis.

The usual way to deal with such data
is the following: The entire separation range under consideration
(from 160\,nm to 750\,nm) is divided into partial subintervals of
length $2\Delta z=1.2\,$nm each. All points from each set of
measurements are plotted together on this interval and, as a
result, each subinterval $j$ contains a group of a few points
$n_j$ (in our
case from 3 to 13). Inside each subinterval all points can be
considered as equivalent because the value of the absolute
separation is distributed uniformly within the limits of
$2\Delta z$.

We note that the experimental points used should be checked for the
presence of so called ``outlying'' results. For this purpose inside
of each subinterval $j$ it is necessary to consider the quantity
\begin{equation}
T_j=\frac{1}{s_{P_{j}}}\mbox{max}|P_{j,i}-\bar{P}_j|,
\label{e6}
\end{equation}
\noindent where $P_{j,i}$ is the value of the Casimir pressure at
point number $i$ of the subinterval $j$ ($1\leq i\leq n_j$,
maximum is taken with respect to $i$). The mean and the
variance of the pressure are respectively defined as
\begin{equation}
\bar{P}_j=\frac{1}{n_j}\sum\limits_{i=1}^{n_j}P_{j,i},
\qquad
s_{P_{j}}^2=\frac{1}{n_j-1}\sum\limits_{i=1}^{n_j}(P_{j,i}-
\bar{P}_j)^2.
\label{e7}
\end{equation}
\noindent If the inequality $T_j > T_{n_{j},1-\beta}$ is satisfied,
where  $T_{n_{j},1-\beta}$ are tabulated quantities,
then the subinterval $j$ contains an outlying result,
which should be rejected with a confidence probability
$\beta$ \cite{32,33}.

The 15 available sets of measurements were analyzed using
this criterion and one set of measurements was found to be
outlying. The subintervals containing points of this set satisfy
the above inequality with high probability $\beta$. As an
example, at separations 170\,nm, 174\,nm, 180\,nm and 250\,nm the
probabilities that the points of the rejected set are outlying are
80\%, 98\%, 95\%, and 98\%, respectively. For this reason in the
subsequent analysis we only consider data from 14 sets of
measurements which were confirmed not to be outlying.

Direct calculation shows that data from these 14 measurement
sets consist of groups (belonging to the neighboring subintervals)
which are uniform in mean values. This implies that the mean
Casimir pressures $\bar{P}_j$ change smoothly in going from group
(subinterval) $j$ to group (subinterval) $j+1$. The variance,
however, turned out not to be uniform in going from one
subinterval to another. For this case we have
utilized the theory of repeated measurements to implement
a special procedure
\cite{34,34a}. At each separation distance $z_0$, in order to find
the uniform variance of a mean, it is necessary to consider not
only one subinterval (which covers the value of $z_0$) but also
several neighboring subintervals to the left and to the right of
$z_0$ (in our case usually 4 or 5) as well. The specific
number of such subintervals $N$ is determined from the requirement
that the variance of the mean computed over several subintervals be
uniform in going to the next set of subintervals.

Mathematically the confidence interval at some fixed separation $z_0$
is found in the following way. We calculate the variance of the mean
of the Casimir pressure in each subinterval $j$ ($1\leq j\leq N$)
 chosen around $z_0$,
\begin{equation}
s_{\bar{P}_{j}}^2=\frac{1}{n_j(n_j-1)}\sum\limits_{i=1}^{n_j}(P_{j,i}-
\bar{P}_j)^2=\frac{s_{P_{j}}^2}{n_j}.
\label{e8}
\end{equation}
\noindent
Then the variance of the mean at a point $z_0$ is found via the equation
\cite{34,34a}
\begin{equation}
s_{\bar{P}}^2(z_0)=\mbox{max}\left[N
\sum\limits_{j=1}^{N}\lambda_j^2s_{\bar{P}_j}^2\right],
\label{e9}
\end{equation}
\noindent
where $\lambda_j$ are the influence coefficients. The maximum is taken over
two sets of influence coefficients, $\lambda_j=1/N$, and
\begin{equation}
\lambda_j=\frac{1}{c_j\sum\limits_{k=1}^{N}c_k^{-1}},
\label{e10}
\end{equation}
\noindent
with the constants $c_k$ determined from
\[
s_{\bar{P}_1}^2:s_{\bar{P}_2}^2:\ldots :s_{\bar{P}_N}^2=
c_1:c_2:\ldots :c_N.
\]
\noindent
The confidence interval at a confidence probability $\beta$
\begin{equation}
\left[\bar{P}_j(z_0)-\Delta^{\! {\rm rand}}P^{\rm expt}(z_0),\,
\bar{P}_j(z_0)+\Delta^{\! {\rm rand}}P^{\rm expt}(z_0)\right]
\label{e11}
\end{equation}
\noindent
can then be found where
\begin{equation}
\Delta^{\! {\rm rand}}P^{\rm expt}(z_0)=s_{\bar{P}}(z_0)
t_{(1+\beta)/2}(\mbox{min}\,n_j-1).
\label{e12}
\end{equation}
\noindent
Here $t_p(f)$ is obtained from the tabulated values for the Student's $t$
distribition (see, e.g.,  \cite{35,36}).

The calculated results for $\Delta^{\!{\rm rand}}P^{\rm expt}(z)$ over
the entire interval (160--750)\,nm at $\beta=0.95$ confidence are shown
in Fig.~9a by the long-dashed line. As is seen from the figure, as
the separation decreases the random error rapidly increases.
This is a consequence of Eqs.~(\ref{e8}) and (\ref{e12}) and the decrease
of $n_j$ for subintervals $j$ at shortest separations.

\subsection{Systematic errors}
We turn now to a discussion of systematic errors in
the Casimir pressure measurements. By convention we call
two of the errors in our experiment
systematic which are in fact not constant, but can vary
within certain limits, and can be described by  random
quantities with a uniform distribution. The first of these is the
sphere radius with the absolute error $\Delta R=0.2\,\mu$m, and the
second is the angular resonant frequency of the oscillator
$\omega_r$ with the absolute error $\Delta\omega_r=2\pi\times
6\,$mHz (Section 3.2).

The determination of the Casimir pressure from Eqs.~(\ref{approx})
and (\ref{e3a}) is in fact an indirect measurement.
However, the error of  $P^{\rm expt}(z)$ should be
expressed in terms of the errors of the directly measured quantities.
In our case $P^{\rm expt}(z)$ is the ratio of two directly
measured quantities,
and for this reason it is convenient to use  the relative
errors (which we denote by  $\delta$), rather than
the absolute errors. Eq.~(\ref{e5}) (written
in terms of the relative errors) can then be used once more \cite{32}
leading to
\begin{equation}
\delta^{\rm syst}P^{\rm expt}(z)=\mbox{min}\left[(\delta\omega_r+\delta R),
1.1\sqrt{(\delta\omega_r)^2+(\delta R)^2}\right],
\label{e12a}
\end{equation}
\noindent where the 95\% confidence level  has been chosen.
Recall that the
quantities $\omega_o$ and $b^2/(I\omega_o^2)$ are determined so
precisely that their uncertainties do not contribute to the
overall error. It should be noted that the first term on the
right-hand side of Eq.~(\ref{e12a}) determines the total result at
large separations of about 450\,nm or more. At small separations
$\delta^{\rm syst}P^{\rm expt}$ is given by the second term  on the
right-hand side of Eq.~(\ref{e12a}).

The absolute systematic error in the Casimir pressure measurements
\hfill \\
$\Delta^{\!{\rm syst}}P^{\rm expt}(z)=
|P^{\rm expt}(z)|\delta^{\rm syst}P^{\rm expt}(z)$,
computed from Eq.~(\ref{e12a}), is shown in Fig.~9a by the
short-dashed line. It is seen that at separations $z > 350\,$nm
the systematic error is slightly greater than the random error. At
the shortest separations the systematic error is negligible.

\subsection{Total experimental error}
To obtain the total experimental error in the Casimir pressure
measurements, it is necessary to combine the random and systematic
errors which are described by different distributions (normal or
Student and uniform in our case). In fact, there are different
methods  in the literature for combining random and systematic errors
\cite{32}. A convenient method in practical applications
\cite{36a}, is based on consideration of the ratio
\begin{equation}
r(z)=\frac{\Delta^{\!{\rm syst}}P^{\rm expt}(z)}{s_{\bar{P}}(z)}.
\label{e13}
\end{equation}
\noindent
At distances where the inequality $r(z) < 0.8$ is satisfied, the
systematic error can be neglected  and the total error of
the Casimir pressure measurements at 95\% confidence level is
given by
\begin{equation}
\Delta^{\!{\rm tot}}P^{\rm expt}(z)=
\Delta^{\!{\rm rand}}P^{\rm expt}(z).
\label{e14}
\end{equation}
\noindent At distances where $r(z) > 8$, the random error can be
neglected and the total error at the same 95\% confidence level is
\begin{equation}
\Delta^{\!{\rm tot}}P^{\rm expt}(z)=
\Delta^{\!{\rm syst}}P^{\rm expt}(z).
\label{e15}
\end{equation}
\noindent
In the intermediate region $0.8\leq r(z)\leq 8$, it is recommended
\cite{36a} that one  use the expression
\begin{equation}
\Delta^{\!{\rm tot}}P^{\rm expt}(z)=
k_{\beta}(r)\left[\Delta^{\!{\rm rand}}P^{\rm expt}(z)
+\Delta^{\!{\rm syst}}P^{\rm expt}(z)\right].
\label{e16}
\end{equation}
\noindent At 95\% confidence level the tabulated coefficient
$k_{\beta}(r)$ varies between
0.71 and 0.81. (For practical purposes
it suffices to use $k_{\beta}(r)=0.8$ \cite{36a}.)

The total error on the measurements of the Casimir pressure
in the separation region (200--750)\,nm is shown in
Fig.~9b by the solid line. For the sake of convenience,
the random and systematic errors are shown
in the same figure by the long-dashed and
short-dashed lines, respectively, on a larger scale than in
Fig.~9a. At separations (160--200)\,nm the total error
coincides with the random one, i.e. is given by the long-dashed
line in Fig.~9a. As can be seen in Fig.~9b, at separations $z\geq
400\,$nm the total experimental error is almost constant and
varies in the interval (0.43,\,0.41)\,mPa.

In Fig.~10, the total relative experimental error
\[
\delta^{\rm tot}P^{\rm expt}(z)=
\Delta^{\!{\rm tot}}P^{\rm expt}(z)/|P^{\rm expt}(z)|
\]
is plotted versus separation. It is seen that the total relative error
is almost constant (varies between 0.55\% and 0.60\%) in a wide
separation region from 170\,nm to 300\,nm.
This contrasts with previous
experiments on the Casimir force, where the smallest relative error
(1.75\% at 95\% confidence \cite{20} in the most precise and accurate
AFM experiment \cite{6}) was achieved only
at the shortest separation.

\section{Different theoretical approaches to the Casimir pressure
between real metals at nonzero temperature}

\subsection{Lifshitz formula combined with the surface impedance or with
the plasma model}

The dynamic
determination of the Casimir pressure, as described in Section 3.2,
was carried out using a Au-coated sphere over a Au-coated plate.
Since the thicknesses of both metal coatings were greater than the
plasma wavelength of Au, $\lambda_p=137\,$nm, one can calculate
the Casimir pressure as if the sphere and plate were composed of
solid Au \cite{12}. In the case of two infinite plates (the correction
due to finite sizes of the plates is negligible, see Section 6) the
result at temperature $T$ is given by the Lifshitz formula
\cite{21,22,23}
\begin{eqnarray}
&&
P(z)=-\frac{k_BT}{\pi}\sum\limits_{l=0}^{\infty}{\vphantom{\sum}}^{\prime}
\int_0^{\infty}k_{\bot}dk_{\bot}q_l
\label{e17} \\
&&
\phantom{aaa}
\times\left\{\left[r_{\|}^{-2}(\xi_l,k_{\bot})e^{2q_lz}-1\right]^{-1}+
\left[r_{\bot}^{-2}(\xi_l,k_{\bot})e^{2q_lz}-1\right]^{-1}\right\}.
\nonumber
\end{eqnarray}
\noindent Here {\boldmath{$k$}}${}_{\bot}$ is the wave vector
component in the plane of the plates,
$k_{\bot}=|\mbox{\boldmath{$k$}}_{\bot}|$,
$q_l^2=k_{\bot}^2+\xi_l^2/c^2$, $\xi_l=2\pi k_BTl/\hbar$ are the
Matsubara frequencies, $k_B$ is the Boltzmann constant,
and $r_{\|,\bot}$ are the reflection
coefficients for two independent polarization states computed
for the imaginary frequencies $\omega_l=i\xi_l$. The prime in
Eq.~(\ref{e17}) refers to the inclusion of a factor 1/2 for
the term with $l=0$.

Eq.~(\ref{e17}) has been derived in the framework of many
different formalisms,
and can hence be considered as the firmly established foundation
for the description of the van der Waals and Casimir pressures
between real materials with finite conductivity at nonzero
temperature. However, the explicit expressions for the reflection
coefficients $r_{\|,\bot}$ are less certain.
 Traditionally, following Lifshitz \cite{21,22,23}, they were
expressed in terms of the dielectric permittivity
$\varepsilon(\omega)$
\begin{equation}
r_{\|,L}^{-2}(\xi_l,k_{\bot})=\left[
\frac{k_l+\varepsilon(i\xi_l)q_l}{k_l-\varepsilon(i\xi_l)q_l}\right]^2,
\quad
r_{\bot,L}^{-2}(\xi_l,k_{\bot})=\left[
\frac{k_l+q_l}{k_l-q_l}\right]^2,
\label{e18}
\end{equation}
\noindent
where $k_l^2=k_{\bot}^2+\varepsilon(i\xi_l)\xi_l^2/c^2$.
For real metals at
nonzero temperature and separations greater than the plasma
wavelength the reflection coefficients can also be represented
in terms of the surface impedance $Z$ \cite{24}
\begin{equation}
r_{\|}^{-2}(\xi_l,k_{\bot})=\left[
\frac{Z(i\xi_l)\xi_l+cq_l}{Z(i\xi_l)\xi_l-cq_l}\right]^2,
\quad
r_{\bot}^{-2}(\xi_l,k_{\bot})=\left[
\frac{Z(i\xi_l)cq_l+\xi_l}{Z(i\xi_l)cq_l-\xi_l}\right]^2.
\label{e19}
\end{equation}
\noindent
At $T=0$ the Lifshitz formula with reflection coefficients given
by Eq.~(\ref{e19}) has long been known \cite{37}. On the real
frequency axis, the surface-impedance reflection coefficients,
Eq.~(\ref{e19}), are commonly used in the analysis of optical
properties of metals \cite{38}. We note that Eq.~(\ref{e19})
contains the Leontovich impedance which does not depend on
polarization or transverse momentum \cite{28}.
If the dielectric permittivity depends only on frequency (as
admitted in the Lifshitz theory) the Leontovich impedance is
given by $Z(\omega)=1/\sqrt{\varepsilon(\omega)}$.
However, the Leontovich impedance and corresponding
boundary conditions on the surface of metal are more general since
they still hold, for
instance, in the frequency domain of the anomalous skin effect
where it is impossible to describe the metal in terms of
$\varepsilon(\omega)$ due to
the spatial nonuniformity of the electromagnetic
field \cite{38}.

It is important to note that expressions (\ref{e18}) and (\ref{e19}) are not
equivalent. They would only become equivalent if  the
Leontovich impedance $Z(\omega)$ in (\ref{e19})
is replaced by the so-called exact impedances
which depend on both polarization and transverse momentum,
\begin{equation}
Z_{\|}(\omega,k_{\bot})=\frac{1}{\omega\varepsilon(\omega)}
\sqrt{\omega^2\varepsilon(\omega)-c^2k_{\bot}^2},
\quad
Z_{\bot}(\omega,k_{\bot})=
\frac{\omega}{\sqrt{\omega^2\varepsilon(\omega)-c^2k_{\bot}^2}},
\label{eI}
\end{equation}
\noindent
where  $\omega$ and $k_{\bot}$ are independent and
not constrained by the mass-shell equation
 $k_{\bot}^2+k_3^2=\omega^2/c^2$
valid for real photons
\cite{28} ($k_3$ is the wave vector component perpendicular
to the plane of the plates). 
Within the fourth (impedance) approach to
the thermal Casimir force it is postulated that the reflection
properties of virtual photons on a metal boundary are the
same as  for real photons, i.e., $\omega$ and $k_{\bot}$ in
Eq.~(\ref{eI}) must be constrained by the mass-shell equation. 
Under this additional condition it then follows that the reflection
coefficients calculated with the exact impedances (\ref{eI}) 
coincide precisely with (\ref{e19}),
expressed in terms of the Leontovich impedance, at zero
frequency and are approximately equal to (\ref{e19}) with a very
high accuracy at all other frequencies (see \cite{28} for more
details). The use of reflection coefficients (\ref{e19}) removes
the contradictions with thermodynamics inherent in the first
approach to the thermal Casimir force
\cite{25,54,55,38a,38b}
which uses the coefficients (\ref{e18}) in combination with the
Drude dielectric function (see Introduction).

Within the experimental separations (160--750)\,nm the
characteristic angular frequency of the Casimir effect
$\xi_c=c/(2z)$ lies within the region of infrared optics.
In the fourth (impedance) approach (see Introduction)
the following calculation procedure is used to compute the
values of the Casimir pressure. First, using the tabulated
optical  data for Au \cite{39}, the dielectric permittivity along
the imaginary frequency axis is computed
\begin{equation}
\varepsilon(i\xi)=1+\frac{2}{\pi}\int_{0}^{\infty}
\frac{\omega\mbox{Im}\varepsilon(\omega)}{\omega^2+\xi^2}d\omega.
\label{e20}
\end{equation}
\noindent
Im$\varepsilon(\omega)$ is found from the tabulated complex refractive
index extending for Au from 0.125\,eV to 10000\,eV \cite{39}.
To calculate $\varepsilon$ at all contributing Matsubara frequencies
with a sufficient precision, it is necessary to extrapolate the
available tabulated data for frequencies below 0.125\,eV. This is usually
done (see, e.g.,  \cite{40,41}) using the imaginary part of the
Drude dielectric permittivity
\begin{equation}
\varepsilon(\omega)=1-
\frac{\omega_p^2}{\omega(\omega+i\gamma)},
\label{e21}
\end{equation}
\noindent
where $\omega_p$ is the plasma frequency and $\gamma$ is the relaxation
frequency (for Au $\omega_p=9.0\,$eV and $\gamma=0.035\,$eV;
$1\,\mbox{eV}=1.51927\times10^{15}\,$rad/s). In all computations
the following values of fundamental constants, recommended by
the National Institute of Standards and Technology,
were used: $c=2.997924\times 10^{8}\,$m/s,
$\hbar=1.05457\times 10^{-34}\,\mbox{J}\cdot\mbox{s}$,
$k_B=1.38065\times 10^{-23}\,$J/K, and
$e=1.602176\times 10^{-19}\,$C.

Once $\varepsilon(i\xi)$ has been found from Eq.~(\ref{e20}),
the values of the
impedance for the imaginary Matsubara frequencies with $l\geq 1$
are calculated as $Z(i\xi)=1/\sqrt{\varepsilon(i\xi)}$. The
resulting values are only slightly different from those computed
from the analytical representation for the surface impedance in
the free electron plasma model [given by Eq.~(\ref{e21}) with
$\gamma=0$],
\begin{equation}
Z(i\xi)=\frac{\xi}{\sqrt{\omega_p^2+\xi^2}}.
\label{e22}
\end{equation}
\noindent
The reason for minor differences is that the
impedance from Eq.~(\ref{e22}) is purely imaginary when evaluated
on a real frequency axis. Eq.~(\ref{e22}) ignores the small real
part of the impedance in the region of infrared optics
arising, for instance, from electron-electron scattering. This real
part is, of course, contained in the tabulated optical  data.

In the fourth approach \cite{24,41a}, the value of the impedance at zero
Matsubara frequency is obtained by extrapolation from the region
of characteristic frequencies (infrared optics in our case). As a
result, the behavior $Z(i\xi)\approx\xi/\omega_p$ when $\xi\to 0$
is obtained from Eq.~(\ref{e22}), which from
Eq.~(\ref{e19}) leads to
\begin{equation}
r_{\|}^{-2}(0,k_{\bot})=1, \qquad
r_{\bot}^{-2}(0,k_{\bot})=
\left(\frac{ck_{\bot}+\omega_p}{ck_{\bot}-\omega_p}\right)^2.
\label{e23}
\end{equation}
\noindent
Using the resulting values of $Z(i\xi_l)$, the Casimir pressure $P_4(z)$
(using the fourth approach)
was computed from Eqs.~(\ref{e17}), (\ref{e19}), (\ref{e20}),
(\ref{e23}) at all
experimental separations for all 14 sets of measurements
at $T=300\,$K. As an illustration, several results are included in
Table 1 (columns 2--4). In column 2 the contribution of the zero-frequency
term of the Lifshitz formula $P_4^{(l=0)}(z)$
is presented.
Column 3 contains the contribution of all terms
$P_4^{(l\geq 1)}(z)$ with $l\geq 1$
in the Casimir pressure. Column 4 presents the values
of the total pressure $P_4(z)$
(the sum of the contributions from the two previous columns).

It is useful to compare the results, obtained by the impedance
approach, with those from  \cite{29,30}, obtained by the
substitution of the plasma model dielectric function
[Eq.~(\ref{e21}) with $\gamma=0$] into the Lifshitz formula
(\ref{e17}) with reflection coefficients (\ref{e18})
(the third approach, see Introduction). In this case
at zero Matsubara frequency, one finds in place of Eq.~(\ref{e23})
\begin{equation}
r_{\|,L}^{-2}(0,k_{\bot})=1, \qquad
r_{\bot,L}^{-2}(0,k_{\bot})=
\left(\frac{ck_{\bot}+\sqrt{c^2k_{\bot}^2+\omega_p^2}}{ck_{\bot}-
\sqrt{c^2k_{\bot}^2+\omega_p^2}}\right)^2.
\label{e24}
\end{equation}
\noindent
The calculated results for the contribution of the term
$P_3^{(l=0)}(z)$
with $l=0$ in the Casimir pressure, the terms
$P_3^{(l\geq 1)}(z)$ with $l\geq 1$, and
 the total pressure $P_3(z)$, are presented in columns 5,\,6,\,7 of Table
1, respectively. In column 8, for comparison, the contribution of
all terms with $l\geq 1$ is calculated using  the tabulated
optical data and Lifshitz reflection coefficients,
Eq.~(\ref{e18}), expressed in terms of the dielectric
permittivity. This column should be compared with column 3 where
the tabulated optical data were used in combination with the
impedance reflection coefficients from Eq.~(\ref{e19}).

As can be seen from Table 1 (columns 4 and 7), at the shortest
separation $z=160\,$nm the plasma model gives smaller
magnitudes of the Casimir pressure than the impedance approach
by approximately 2.6\%.
The same holds at zero temperature \cite{40} where the
situation is quite clear and there are no conflicts among
the different approaches. This
underestimation decreases to only 0.7\% with increasing
separation until $z=250\,$nm. At greater separations the
differences among the data in columns 4 and 7 quickly decrease.
The above differences at short separations are explained by
the fact that the plasma model does not include the internal
photoelectric effect (interband transitions) and other processes
taken into account by the tabulated optical
 data. On the other hand, if we compare columns 3 and 8 of Table 1,
it is seen that the differences are less than 0.42\% (the greatest
difference being at the shortest separation $z=160\,$nm).
It follows that the impedance reflection coefficients from
Eq.~(\ref{e19}) lead to practically the same results as the
Lifshitz coefficients given by Eq.~(\ref{e18}).

\subsection{Lifshitz formula with zero-frequency term determined
by the Drude model or by special prescription}

As was noted in the Introduction, the first and second
 alternative approaches
to the description of the thermal Casimir force between real metals
lead to predictions which differ
from those in Section 5.1. Here we compute the
Casimir pressure in the framework of these approaches
as they were formulated by their authors.

We start with the first approach \cite{25} in which
the Lifshitz formula (\ref{e17}) and the reflection coefficients
(\ref{e18}) were used to calculate the Casimir pressure between Au
plates, taking into account the effects of nonzero temperature and
finite conductivity. The values of the dielectric permittivity at
 imaginary Matsubara frequencies $\varepsilon(i\xi_l)$ were
obtained in the same way as was described in Section 5.1. In other words,
tabulated optical  data, extrapolated in the low frequency range
by the imaginary part of the Drude dielectric function, were used
to obtain $\varepsilon(i\xi_l)$ by means of the dispersion relation
(\ref{e20}). Up to this point the approach of  \cite{25}
almost coincides with the impedance approach.
(As was demonstrated in Section 5.1,
column 8 of Table~1, the use of the Lifshitz reflection
coefficients instead of the impedance coefficients leads to only minor
differences at all Matsubara frequencies with $l\geq 1$.)
The major difference in the approach in  \cite{25} is the value used
 for the zero-frequency term of the Lifshitz formula. Instead
of Eq.~(\ref{e23}), the reflection coefficients at zero Matsubara
frequency were obtained in  \cite{25} by the substitution of the
Drude dielectric function (\ref{e21}) into Eq.~(\ref{e18}) with
a result
\begin{equation}
r_{\|,L}^2(0,k_{\bot})=1, \qquad
r_{\bot,L}^2(0,k_{\bot})=0.
\label{e39}
\end{equation}
\noindent
The approach of  \cite{25}
was supported in  \cite{54,55}
(see also \cite{38a,38b}).

Using the calculational procedure of  \cite{25,54,55}, the
Casimir pressure $P_1(z)$ obtained from the first approach
was computed at all experimental
separations at $T=300\,$K. Several results are presented in Table~2.
In columns 2,\,3 the zero-frequency term $P_1^{(l=0)}(z)$ and
the total Casimir pressure $P_1(z)$ are given, respectively.
(Recall that the contributions of the Matsubara frequencies with
$l\geq 1$ were already included in column 8 of Table 1.)

We next turn our attention to the second alternative approach to the
thermal Casimir force suggested in  \cite{27}.
This approach also
used the Lifshitz formula (\ref{e17}) with the reflection coefficients
(\ref{e18}) to calculate the Casimir pressure between the plates made
of real metals at nonzero temperature. The dielectric permittivity
$\varepsilon(i\xi_l)$ was found in the same way as in  \cite{25}.
The major difference between
the approach \cite{27} and those used in
Section 5.1, or  in  \cite{25}, is a new
alternative for the zero-frequency term of the Lifshitz formula.
Instead of Eqs.~(\ref{e23}), (\ref{e24}) or (\ref{e39}),
it was postulated that for real metals
both reflection coefficients at zero Matsubara frequency
 should be the same as for ideal metals, i.e.
\begin{equation}
r_{\|}^2(0,k_{\bot})=1, \qquad
r_{\bot}^2(0,k_{\bot})=1.
\label{e41}
\end{equation}
\noindent
Using this postulate, the Casimir pressure $P_2(z)$
obtained from the second approach was computed from
Eqs.{\ }(\ref{e17}), (\ref{e18}) at all experimental separations at
$T=300\,$K.  As an example, several
calculated results are presented in columns 4,\,5 in Table~2,
where the zero-frequency term $P_2^{(l=0)}(z)$
and the total Casimir pressure $P_2(z)$, obtained from
Eqs.~(\ref{e41}) and (\ref{e17}),
are given, respectively.

Comparing column 4 of Table 1 with column 3 of Table 2, we conclude
that the first approach leads to smaller magnitudes of the Casimir pressure
than the impedance approach at all separations. As an illustration, the
differences of the theoretical predictions given by these two
approaches change from 2.4\% of the Casimir pressure at
$z=200\,$nm to 7.4\% at $z=500\,$nm. On the contrary, if we compare
column 4 of Table 1 with column 5 of Table 2, we conclude that
the second approach predicts larger magnitudes for the Casimir pressure
than does the impedance approach. The differences of the theoretical
predictions given by these two approaches change slowly
from 2.5\% of the Casimir pressure at $z=200\,$nm to 2.2\%
at $z=500\,$nm.
It should be emphasized that the calculated results in Table 2
are in agreement with those obtained
by the authors of \cite{25,54,55,38a,38b} (columns 2,\,\,3)
and \cite{27} (columns 4,\,\,5), where 
the first and second approaches,
respectively, were developed.

\subsection{Variation of optical tabulated data and effects
of spatial nonlocality}

To ensure that the theoretical approaches which
use  tabulated optical
  data for the complex refractive index, lead to
sufficiently precise results, one should check that small
sample-to-sample variations of the data due to, for example, size of
the grains and impurities are not problematic. This was
verified in  \cite{20} for the
experiment of  \cite{6}. It was found that at $z=62\,$nm the
possible variations of the optical data due to grain size lead to a
0.8\% decrease of reflectivity, and to a decrease of less than 0.5\%
in the Casimir force. In our case the sizes of grains vary between
25\,nm and 300\,nm (with a mean value of about 150\,nm), and the
characteristic wavelength at $z=160\,$nm is $\lambda_c=2\pi
c/\xi_c=4\pi z\approx 2010\,$nm. According to  \cite{42},
even for grains as small as 45\,nm in size (the largest studied in that paper)
the reflectivity at $\lambda\sim 2000\,$nm is only 0.6\% less than
that calculated from tabulated data. Taking into account
that the mean size of crystallite grains in our case is several
times larger than 45\,nm, we can conclude that the variations of the
optical data may lead to a decrease in the
magnitude of the Casimir pressure much lower than 0.5\%,
even at the shortest separation considered.

One more effect to discuss, related to the finite conductivity
corrections, is the probable contribution from  the effects of
spatial nonlocality (i.e., wave vector dependence of the
dielectric function). These effects influence the Casimir force
magnitude in the region of the anomalous skin effect at large
separations, $z>2.36\,\mu$m for Au \cite{24}, a region not relevant
for our experiment. Another, short-separation, region where
nonlocality may contribute to the van der Waals force, is
$z<\lambda_p/(4\pi)\approx 10.9\,$nm which corresponds to
characteristic frequencies $\xi_c>\omega_p$. At such high
frequencies the spatial dispersion modifies the frequencies
of surface plasmons. In our experiment,
however, the effects of nonlocality do not
contribute to the Casimir pressure.
This can be understood by noting that the
largest characteristic frequency here, calculated at $z=160\,$nm,
is $\xi_c=9.4\times 10^{14}\,$rad/s, which is almost 15 times smaller
than $\omega_p=1.37\times 10^{16}\,$rad/s. A direct calculation
shows that the Matsubara frequencies
$\xi_0,\,\xi_1,\ldots,\xi_{35}$ contribute 99.96\% of the total
Casimir pressure at $z=160\,$nm, and even $\xi_{35}=8.61\times
10^{15}\,\mbox{rad/s}<\omega_p$.
We note in passing that the contribution of the nonlocal effects
for Au of about
2\% at a separation $z=\lambda_p=137\,$nm, claimed in
\cite{43}, is in error, as was recognized by the authors in a recent
Erratum \cite{43a}.
 (In fact the formalism of this reference
leads to a correction of only 0.2\% due to spatial nonlocality at
$z=\lambda_p$; the confusion arises from a notation
$L/\lambda_p$ on the $x$-axis in Fig.~2 of  \cite{43} instead
of $2\pi L/\lambda_p$.) Furthermore, in  \cite{43}
a spatially nonlocal dielectric function was used in the
frequency region of infrared optics where it is in fact local
\cite{38,44}. As was specifically demonstrated in detail in
 \cite{45}, the
short-separation effects of spatial nonlocality connected with
the surface-plasmon charge fluctuations contribute considerably to
the force at separations of the order of $v_F/\omega_p\sim
1\,${\AA}, where $v_F$ is the Fermi velocity. It was explicitly
proved in  \cite{45} that for much larger distances $z\gg
v_F/\omega_p$ (in fact for $z$ larger than several nanometers) the
usual local Lifshitz formula is applicable.
This is in agreement with our
conclusion that at separations $z\geq 160\,$nm there is no
noticeable contribution from the effects of nonlocality
to the Casimir pressure.

\subsection{Casimir pressures taking account of surface roughness}
At separations below 1\,$\mu$m surface roughness corrections
may contribute from a fraction of a percent up to 20\% of the measured
force in different experiments \cite{12}. For this reason they
should be carefully taken into account in precision computations of
the Casimir pressure.
The contribution of roughness is calculated with the help of
the AFM characterization of both surfaces (see Section 2.2).
As is seen in Figs.~2 and 3, the roughness is mostly represented by
stochastically distributed distortions with a characteristic
lateral size of approximately $l\approx(500-600)\,$nm. This size is larger
by a factor of $\approx 3$
than $z$ at the shortest separations (160--200)\,nm,
where the roughness corrections may
give a noticeable contribution to the Casimir
pressure in our experiment. The influence of roughness can be calculated
by the additive method of geometrical averaging \cite{12}. As shown in
 \cite{47}, under the condition $z<l$ the additive method gives
approximately the same results as more detailed methods. Moreover,
the correction due to nonadditivity can be simply estimated.

There are two modifications to the additive method. The
first one is based on the use of the Casimir pressures $P_i(z)$,
found in the above four theoretical approaches
as calculated in Sections 5.1, 5.2. Let us first determine the
zero roughness levels $H_o^{p,s}$ on a plate and a sphere, respectively,
relative to which the mean values of the functions, describing
roughness, are zero (note that the experimental separations are
measured between the zero roughness levels \cite{12}):
\begin{equation}
\sum\limits_{i=1}^{K}\left(H_o^{p,s}-h_i^{p,s}\right)
v_i^{p,s}=0.
\label{e25}
\end{equation}
Here the quantities $v_i^{p,s}$ and $h_i^{p,s}$ are taken from Figs.~4a
and 4b for the plate and the sphere, respectively
($K$ was defined in Section 2.2).
To find the Casimir pressure taking into account all three corrections,
the values $P(z)$ should be geometrically averaged over all possible
separations between the rough surfaces, weighted with the probability
of each separation \cite{3,11,12}
\begin{equation}
P_i^{\rm theor}(z)=\sum\limits_{k=1}^{105}\sum\limits_{j=1}^{112}
v_k^pv_j^sP_i\left(z+H_o^p+H_o^s-h_k^p-h_j^s\right).
\label{e26}
\end{equation}
\noindent
Note that this expression does not reduce to a simple multiplication of
the correction factors due to nonzero temperature and finite conductivity
on the one hand, and due to surface roughness on the other, but takes into
account their combined (nonmultiplicative) effect.

The second one is a simpler additive method. It was used
for calculating the corrections due to
stochastic surface roughness in  \cite{48}.
According to this method, the Casimir pressure between two
ideal metal parallel plates
covered by stochastic roughness with
variances $\delta_{st}^p$ and $\delta_{st}^s$ is given by
\begin{equation}
P_r(z)=\eta_rP_o(z), \quad
\eta_r=1+10\left[\left(\frac{\delta_{st}^p}{z}\right)^2+
\left(\frac{\delta_{st}^s}{z}\right)^2\right],
\label{e27}
\end{equation}
\noindent
where $P_o(z)=-\pi^2\hbar c/(240z^4)$ is the Casimir pressure between
perfectly shaped parallel plates. The variances
describing the stochastic roughness are found from the formula
\begin{equation}
\left(\delta_{st}^{p,s}\right)^2=\sum\limits_{i=1}^{K}
\left(H_o^{p,s}-h_i^{p,s}\right)^2v_i^{p,s}.
\label{e28}
\end{equation}
\noindent
Using data from Figs.~4a,\,4b, one obtains the values for variances
$\delta_{st}^p=4.06\,$nm and $\delta_{st}^s=1.91\,$nm.
The smallness of these variances
justifies the neglect of the fourth order terms on the right-hand side
of Eq.~(\ref{e27}) (the fourth order contributes less than 0.01\% even
at the shortest separation $z=160\,$nm).
We can now obtain the complete theoretical
result, including the corrections due to surface roughness, nonzero
temperature and finite conductivity,  by the use of
a multiplicative procedure
\begin{equation}
P_i^{\rm theor}(z)=\eta_r P_i(z).
\label{e29}
\end{equation}

It is interesting to compare the theoretical results for the Casimir
pressure taking into account all corrections
contained in  Eq.~(\ref{e26}), and the simpler expression in
Eq.~(\ref{e29}). As an example, at the shortest separation $z=160\,$nm,
Eq.~(\ref{e26}) gives $P_4^{\rm theor}=-1.1515\,$Pa and
Eq.~(\ref{e29}) results in $P_4^{\rm theor}=-1.1531\,$Pa, a difference
only 0.14\% of the Casimir pressure. At $z=200\,$nm the complete Casimir
pressures are $P_4^{\rm theor}=-0.51143\,$Pa and
$P_4^{\rm theor}=-0.51185\,$Pa
computed from Eqs.~(\ref{e26}) and (\ref{e29}), respectively. In this case
the difference is even smaller: 0.08\% of the Casimir pressure. For larger
separations the role of the nonmultiplicative effects decreases and
one may use Eq.~(\ref{e29}) to calculate the theoretical values of the
Casimir pressure with sufficiently small uncertainty. The same relative
sizes of roughness corrections are obtained for the other theoretical
approaches.

Note that the role of the surface roughness in the improved experiment
is rather modest. Using the data of Tables 1 and 2  we find
that at $z=160\,$nm roughness
contributes 0.65\%,
and at a separation $z=200\,$nm only 0.42\% of the Casimir pressure.

It is important to bear in mind that both Eqs.~(\ref{e26})
and (\ref{e29}) are
based on the approximation and do not take
into account diffraction-type effects arising in the case of
periodic roughness with small periods $l<z$ \cite{47}, or stochastic
roughness with small correlation length \cite{49}.
To obtain an estimate for the upper limit of the contribution of
diffraction-type effects we follow  \cite{20}. For this purpose,
the bound on the roughness correlation length is found to be
$l_{corr}\geq 600\,$nm, where 600\,nm is the lateral size of the largest
cluster on the surface. We next consider periodic roughness with
a period $l_{corr}=600\,$nm using the fact that diffraction-type
effects are greater for a periodic function with a period $l_{corr}$
than for a random function with a correlation length $l_{corr}$. The
diffraction-type effects for a periodic function can be computed
in the framework of the functional approach of  \cite{47}.
At separations $z=160\,$nm and 300\,nm one obtains
$z/l_{corr}=0.27$ and 0.50, respectively. The coefficient
$\tilde{c}_{corr}$ in the expression
\begin{equation}
\eta_r^{corr}=1+10\tilde{c}_{corr}\left[\left(
\frac{\delta_{st}^p}{z}\right)^2+\left(
\frac{\delta_{st}^s}{z}\right)^2\right],
\label{e30}
\end{equation}
\noindent
taking diffraction-type effects into account, is related to that
 plotted in Fig.~2 of  \cite{47} by the equation
\begin{equation}
\tilde{c}_{corr}=c_{corr}-\frac{1}{5}z
\frac{\partial c_{corr}}{\partial z}.
\label{e31}
\end{equation}
\noindent
(In  \cite{47} the results for the Casimir energy between plates
are presented, not those for the pressure.) Using the data of Fig.~2
from paper \cite{47} and Eq.~(\ref{e31}), we find
$\tilde{c}_{corr}\approx 1$
and $\tilde{c}_{corr}\approx 1.18$ at separations $z=160\,$nm and
$z=300\,$nm, respectively. Hence at $z=160\,$nm there is no
noticeable contribution from correlation effects to the roughness
correction. From Eqs.~(\ref{e27}) and (\ref{e30}) it
follows that at $z=300\,$nm  $\eta_r=1.0022$ and
$\eta_r^{corr}=1.0026$, leading to a
contribution from correlation effects to the roughness correction of
less than 0.04\%. Because of this, the correlation effects of roughness
in the experiment under consideration can be ignored.

\subsection{Contribution of patch potentials}
In the experimental configuration
described in Sections 2,\,3, the electrostatic
force due to the residual potential difference between the plate
and the sphere is negligible. It is conceivable, however, that
spatial variations of the surface potentials due to grains of
polycrystalline metal film (the so called ``patch potentials'')
simulate the Casimir force \cite{50}. Here we use the general
results of  \cite{50} in order to demonstrate that the patch
effect is negligible in our experiment.
According to  \cite{50},
 for a configuration of two parallel plates
the electric pressure due to random variations in patch potentials is
given by
\begin{equation}
P^{patch}(z)=-\frac{2\varepsilon_0\sigma_v^2}{k_{\max}^2-k_{\min}^2}
\int_{k_{\min}}^{k_{\max}}\frac{k^3}{\sinh^2 kz}dk.
\label{e32}
\end{equation}
\noindent
Here $\sigma_v$ is the variance of the potential distribution,
and $k_{\max}$ ($k_{\min}$) are the magnitudes of the extremal
wavevectors corresponding to minimal (maximal) sizes of grains.
Using the work functions of Au  for different crystallographic
surface orientations
($V_1=5.47\,$eV, $V_2=5.37\,$eV,
and $V_3=5.31\,$eV),
and assuming
equal areas of respective crystallographic planes, we obtain
\begin{equation}
\sigma_v^2=\frac{1}{{2}}
\sum\limits_{i=1}^{3}(V_i-\bar{V})^2
\approx 6528.64\,\mbox{mV}^2.
\label{e33}
\end{equation}

As was mentioned in Section 5.3, the extremal sizes of grains in the Au layers
covering the test bodies are $\lambda_{min}\approx 25\,$nm and
$\lambda_{max}=300\,$nm. This results in $k_{max}=0.251\,\mbox{nm}^{-1}$
and $k_{min}=0.0209\,\mbox{nm}^{-1}$. Substituting these values into
Eq.~(\ref{e32}) yields the additional pressures due to the patch
electric forces. At the shortest separations $z=160\,$nm and $z=170\,$nm
the patch pressures are  $P^{patch}=0.42\,$mPa and 0.25\,mPa,
respectively. In comparison to  the Casimir pressures at the same
separations, the relative contributions of the patch effect to the
pressure are 0.037\% and 0.027\%, respectively, and
further decrease with increasing $z$. It follows that
patch effects do not play any significant role in our Casimir
pressure measurements
by means of a micromachined oscillator.

\section{Metrological analysis of theoretical errors}

As it is evident from the foregoing, the theoretical computation of the
Casimir pressure with all relevant corrections is a rather complicated
procedure requiring a variety of data. Bearing in mind the
eventual comparison of theory
with experiment, it is important to analyze all uncertainties introduced
at different stages of the computations.

According to Eq.~(\ref{e3a}), the derivative of the Casimir force
between a plate and a sphere with respect to separation is related
to the Casimir pressure between two plane plates by the use of
the proximity force theorem. It is well known that this ``theorem'' is
approximate and its relative error $\delta_{pft}P^{\rm theor}$ is less
than $z/R$ \cite{51,52,52a}. This error is separation-dependent and
increases from 0.107\% at a separation $z=160\,$nm to 0.5\% at
$z=750\,$nm (see the short-dashed line in Fig.~11).

The computation of the Casimir pressure taking into account nonzero
temperature and finite conductivity in Sections 5.1, 5.2 was based on the
use of tabulated optical  data for Au (approaches 1, 2 and 4).
The primary source of the theoretical errors is the sample to sample
variation of these data. As was shown in Section 5.3, in the
experiment under consideration the variation of the optical data
leads to an uncertainty in the magnitude of the Casimir pressure which
is much smaller than 0.5\%.
To be conservative, we admit an uncertainty in the computation of
the finite conductivity corrections as large as
$\delta_cP_i^{\rm theor}=0.5$\% over the
entire measurement range.
In fact, this error also includes all differences arising
when one uses the Lifshitz reflection coefficients (\ref{e18})
instead of impedance coefficients (\ref{e19}) and vice versa (see
Section 5.1). The theoretical error $\delta_c$ is shown in Fig.~11
by the long-dashed line.

The other uncertainties discussed above, such as the
contribution of nonlocality
to the effect of finite conductivity (Section 5.3),
the diffraction-type contributions to the effect of surface
roughness (Section 5.4), and the correction to the Casimir pressure due
to the patch potentials (Section 5.5), were shown to be far smaller than
the previous two
discussed above, $\delta_{pft}P_i^{\rm theor}$ and
$\delta_cP_i^{\rm theor}$.
For this reason they can all be neglected. One further correction to
the Casimir pressure, arising from the finite size of plates, is also
negligible. As was shown in  \cite{11}, in the plate-sphere
configuration of a micromachined oscillator this correction is less
than 0.04\% at $z=500\,$nm, becoming smaller as the separation
decreases.  But even this correction does not contribute in our
case of two equivalent parallel plates because its derivative with
respect to separation vanishes. In fact, to obtain the correction
due to the finiteness of the plates, one should consider
corrections of higher
order in $z/R$ for a sphere above a plate. As a result,
at least an extra power of the small parameter $z/R$ appears in the
correction, making it completely insignificant in the error analysis.

Let us now determine the theoretical error resulting from
combining the two major errors $\delta_{pft}P_i^{\rm theor}$ and
$\delta_cP_i^{\rm theor}$.
Both random quantities are described by  the same distribution
law which is close to a uniform distribution. For this reason the
method of  \cite{32}, already used in Eqs.~(\ref{e5}), (\ref{e12a}),
 can be applied once more, giving
\begin{eqnarray}
&&
\delta_oP_i^{\rm theor}(z)=
\mbox{min}\left[\left(\delta_{pft}P_i^{\rm theor}+\delta_cP_i^{\rm theor}
\right),\right.
\label{e34} \\
&&
\phantom{aaaaaaaaaaaaaaaaaaa}\left.
1.1\sqrt{(\delta_{pft}P_i^{\rm theor})^2+(\delta_cP_i^{\rm theor})^2}
\right].
\nonumber
\end{eqnarray}
\noindent
As in Eqs.~(\ref{e5}) and (\ref{e12a}), the 95\% confidence level is
chosen. Note that in this case the minimum is achieved
using the second
term on the right-hand side of Eq.~(\ref{e34}).

Bearing in mind the ensuing comparison of theory with experiment,
it is pertinent to consider the error in the theoretical pressures
resulting from the experimental error in the determination of
separation distances $\Delta z$ \cite{53}. Although this error
is not purely theoretical (theoretical pressures by themselves can
be calculated at any exact separation distance), it is of great
importance when one computes the theoretical pressure at
an experimental point defined with an error $\Delta z$. Using the
main theoretical dependence of the Casimir pressure on the inverse
fourth power of the separation one obtains
$\delta_zP_i^{\rm theor}=4\Delta z/z$, where $\Delta z=0.6\,$nm at 95\%
confidence [see Eq.~(\ref{e5})]. The value of $\delta_zP_i^{\rm theor}$
changes from 1.5\% at $z=160\,$nm to 0.32\% at $z=750\,$nm
(see the dotted line in Fig.~11). Thus, this error is the primary one
(especially at short separations) arising in  the comparison
of theory with experiment.

In  \cite{20} an additional fitting procedure was proposed
in order to decrease the error in the determination of absolute separations
down to $\Delta z=0.15\,$nm. For this purpose, for each set of measurements,
$z$-values of all points were simultaneously changed within the limits
of their joint absolute error in order to minimize the root-mean-square
deviation between theory and experiment. This procedure, which is quite
reasonable at very short separations ($z\geq 60\,$nm in  \cite{20}),
is not applied in our conservative metrological analysis because the
method of least squares is not well-adapted for cases when both the
argument (separation distance) and function (pressure) are determined
with comparable relative errors.

We are now in a position to determine the total theoretical error of the
Casimir pressure computations.
To do this, we combine the errors determined above,
$\delta_oP_i^{\rm theor}$ and $\delta_zP_i^{\rm theor}$.
To use the analogy with Section 4.3,
the error $\delta_oP_i^{\rm theor}$ can be considered as ``random''
(in the sense that it is not described by a uniform distribution),
and the error $\delta_zP_i^{\rm theor}$ can be likened to ``systematic''
(as described by a uniform distribution). We then find ourselves
in the regime of applicability  of Eq.~(\ref{e16}), and at 95\% confidence,
we obtain the final result
\begin{equation}
\delta^{\rm tot}P_i^{\rm theor}(z)=0.8\left[\delta_zP_i^{\rm theor}(z)+
\delta_oP_i^{\rm theor}(z)\right].
\label{e35}
\end{equation}
\noindent
This result is represented by the solid line in Fig.~11.
It is seen that the total theoretical error decreases from 1.65\% to 1\%
when the separation increases from 160\,nm to 380\,nm. With further increase
of separation the total theoretical error decreases more slowly from
1\% to 0.9\%.

\section{Comparison of theory and experiment and stronger constraints
on thermal effects}

\subsection{Measure of agreement between theory and experiment}

In all previous experiments on the Casimir effect
\cite{2,3,4,5,6,7,8,9,10,11,12} the agreement between theory and
experiment was discussed in terms of the root-mean-square
deviation between experimental and theoretical values of the force
over the entire range of separation distances.
In  \cite{7} this approach was
criticized as inadequate when the force decreases rapidly as
the separation distance increases,
although no alternative approach was suggested.
Here we compare experiment and theory with the help of rigorous
metrological criteria, not based on the concept of
root-mean-square deviation.

{}From Sections 4.3 and 6, we have at our disposal two independently
obtained confidence intervals at the same confidence probability of 95\%
(the first for experiment and the second for theory). Both confidence
intervals were determined by the application of a common statistical
procedure, and the resulting errors for both experiment and theory can be
described by random variables with a common distribution law. With
this in mind, we consider the new variable
$\left[P_i^{\rm theor}(z)-P^{\rm expt}(z)\right]$
and determine the total absolute error on this quantity, and hence
the confidence interval at a confidence probability 95\%, using
the same composition law as in Eqs.~(\ref{e5}) and (\ref{e12a}), i.e.
\begin{eqnarray}
&&
\Delta^{\!{\rm tot}}\left[P_i^{\rm theor}(z)-P^{\rm expt}(z)\right]
\phantom{\,}=\mbox{min}\left\{
\left[\Delta^{\!{\rm tot}}P^{\rm expt}(z)+
\Delta^{\!{\rm tot}}P_i^{\rm theor}(z)\right],\right.
\label{e36} \\
&&
\phantom{aaaaaa}\left.
1.1\sqrt{\left[\Delta^{\!{\rm tot}}P^{\rm expt}(z)\right]^2+
\left[\Delta^{\!{\rm tot}}P_i^{\rm theor}(z)\right]^2}\right\}.
\nonumber
\end{eqnarray}
\noindent
Here $\Delta^{\!{\rm tot}}P^{\rm expt}(z)$ is given by the solid
line in Fig.~9b, and
$\Delta^{\!{\rm tot}}P_i^{\rm theor}(z)=
P_i^{\rm theor}\,\delta^{\rm tot}P_i^{\rm theor}$, where
$\delta^{\rm tot}P_i^{\rm theor}$ is shown by the solid line in Fig.~11.
Note that the total absolute theoretical error is almost
independent of the theoretical approach. For this reason we omit the
index $i=1,\,2,\,3,\,4$ in the notation for a confidence interval.
The minimum here is achieved using the second contribution
from the right-hand side of Eq.~(\ref{e36}). We also point out that
to be conservative we use the value $k_{\beta}=1.1$, as would
describe the
composition of two uniform distributions (for other distributions the
value of $k_{\beta}$ can be somewhat smaller).

The confidence interval for the quantity
$[P_i^{\rm theor}(z)-P^{\rm expt}(z)]$ at 95\% confidence probability
is given by
\begin{equation}
\left[-\Delta^{\!{\rm tot}}\left(P^{\rm theor}(z)-
P^{\rm expt}(z)\right),\,
\Delta^{\!{\rm tot}}\left(P^{\rm theor}(z)-
P^{\rm expt}(z)\right)\right].
\label{e37}
\end{equation}
\noindent
The meaning of the confidence interval (\ref{e37}) is that,
if the theory is in agreement with experimental data, the mean value of
$[P_i^{\rm theor}(z)-P^{\rm expt}(z)]$ must belong  to this interval with
a 95\% probability. This criterion can be used for comparison of the
above four theoretical approaches with experiment.

\subsection{Comparison of different theoretical approaches with
experiment}

We start from the approach based on the surface impedance
(the fourth approach).
The solid lines in Fig.~12 exhibit the confidence interval
(\ref{e37}) obtained from Eq.~(\ref{e36}) within the separation ranges
(160--250)nm (a) and (250--750)nm (b). In the same figure the differences
of the theoretical (based on the surface impedance)
and experimental Casimir pressures are plotted for one
typical set of 14 measurements. It is evident that the true
value of the quantity under consideration
$\left[P_4^{\rm theor}-P^{\rm expt}\right]$ (which can only
be achieved for a complete correct theory and an infinite number of
measurements) is zero.
As an example, for the measurement set plotted in
Fig.~12, $\langle P_4^{\rm theor}-P^{\rm expt}\rangle=0.008\,$mPa,
where the averaging
was taken over all 288 points belonging to this set. If only points
at separations $z\geq 250\,$nm are considered (242 points), one obtains
$\langle P_4^{\rm theor}-P^{\rm expt}\rangle=0.09\,$mPa.
These values should be compared
with the half-width of the confidence interval which decreases from
27.7\,mPa at $z=160\,$nm to 0.46\,mPa at $z=750\,$nm. Remarkably, only
9 of 288 points (i.e. 3.1\%) fall outside
the confidence interval in Fig.~12.

In Fig.~13a,b the same confidence interval (\ref{e37}) is shown once more
and the points
$\left[P_4^{\rm theor}-P^{\rm expt}\right]$ are plotted utilizing  all 14
sets of measurements. For all 4066 points in Fig.~13 one has
$\langle P_4^{\rm theor}-P^{\rm expt}\rangle=0.11\,$mPa. If only separations
$z\geq 250\,$nm are considered (3442 points) the result is
$\langle P_4^{\rm theor}-P^{\rm expt}\rangle=0.12\,$mPa.
Although these values are
less than the half-width of the confidence interval, they are not
sufficiently informative. In the case of Fig.~13,
where all points are plotted,
it is much more important to consider the local values of the quantity
$\Delta\bar{P}_i(z)\equiv\langle P_i^{\rm theor}(z)-P^{\rm expt}(z)\rangle$,
where the averaging is done over
the vicinity of a point $z$, in comparison with the local half-width
of the confidence interval. Performing the averaging over the intervals
($z_i-3\,\mbox{nm},z_i+3\,\mbox{nm}$) (so that each contains approximately
 43 points) at different $z=z_i$ one obtains the values for the
mean differences between the theoretical (the fourth approach)
and experimental Casimir pressures
included in column 3 of Table~3. For comparison, the
values of the half-width of a confidence interval are included
in column 2. Columns 4--6 are discussed below.

Comparing columns 2 and 3 of Table~3 we notice that all mean
values of the differences between theoretical and experimental
Casimir pressures are well within the half-width of the
confidence interval. The number of separate points which fall outside
of the confidence interval in Fig.~13 is 207, i.e. 5.09\% of the total
number of points from the fourteen sets of measurements.
This is in accordance with expectations from
the value of the confidence probability.
To conclude, the experimental data of Section 3.2 are proved to be
in excellent agreement with the theoretical computations of the Casimir
pressure in the framework of the impedance approach (see also
 \cite{Lam1,Lam2} arriving to the result in agreement with
this approach).

Quantitatively the agreement between theory and experiment can be
characterized by the quantity
\begin{equation}
\delta^{\rm tot}(z)=
\frac{\Delta^{\!{\rm tot}}
\left[P^{\rm theor}(z)-P^{\rm expt}(z)\right]}{|P_4^{\rm theor}(z)|}.
\label{e38}
\end{equation}
\noindent
Importantly, in the experiment under consideration $\delta^{\rm tot}(z)$
depends only slightly on $z$ in a wide separation range, decreasing
(with the fourth theoretical approach)
from $\delta^{\rm tot}(z)=1.9$\% at $z=170\,$nm to
$\delta^{\rm tot}(z)=1.4$\%  within the
$270\,\mbox{nm}\leq z\leq 370\,$nm interval, and then increasing to
$\delta^{\rm tot}(z)=1.8$\% at $z=420\,$nm. The largest values of
$\delta^{\rm tot}$ at $z<170\,$nm and $z>420\,$nm are
$\delta^{\rm tot}(z)=2.4$\% at $z=160\,$nm and
$\delta^{\rm tot}(z)=13$\% at $z=750\,$nm.
Thus, the Casimir pressure measurement by means of a micromachined
oscillator is
the first Casimir effect experiment where metrological
agreement between theory and experiment
at the 1.5\% level has been achieved at 95\% confidence in a wide
separation region.

The results of a comparison between experiment and the third theoretical
approach, based on the plasma dielectric function,
are presented in column 4 of Table 3. Comparing column 4 with column 2,
one can conclude that the third approach is also consistent with
the experimental
data because all mean values of the Casimir pressures belong to
the confidence interval. At the same time, it is seen
that at the shortest separations $(170,\,250)\,$nm this
approach does not agree with experiment as well as does the impedance
approach. The reasons for this
were discussed in Section 5.1 (see also  \cite{40}).

We next compare the first theoretical approach with the experiment.
In Fig.~14a, we plot the differences of the Casimir pressures
$\left[P_1^{\rm theor}-P^{\rm expt}\right]$
versus separation $z$ (where $P_1^{\rm theor}$ was computed
in the approach of  \cite{25,54,55} as explained above) for all
14 sets of measurements (i.e., in the same way
as in Fig.~13 for the surface impedance approach). The solid line
gives the half-width of the confidence interval
$\Delta^{\!{\rm tot}}\left(P_1^{\rm theor}-P^{\rm expt}\right)\approx
\Delta^{\!{\rm tot}}\left(P_4^{\rm theor}-P^{\rm expt}\right)$  at 95\%
confidence (note that there are practically no points below the
$z$-axis).
As is seen from Fig.~14a, in a wide separation region
$230\,\mbox{nm}\leq z\leq 500\,$nm all points
$\left[P_1^{\rm theor}-P^{\rm expt}\right]$
fall outside the confidence interval. It follows
 that the theoretical prediction of the first approach
is excluded by experiment at 95\% confidence.

In column 5 of Table~3 the mean values of the differences between
theoretical (obtained from the first approach)
and experimental Casimir pressures are presented at different separations
from 170\,nm to 700\,nm using the data
from all 14 sets of measurements.
The comparison with the half-width of the confidence interval (column 2)
shows that at all separations the mean differences of the Casimir
pressures fall outside the confidence interval. It follows that
the first theoretical approach is excluded by the
experiment at 95\% confidence level within an even wider separation region,
$170\,\mbox{nm}\leq z\leq 700\,$nm.

In fact, the above confidence interval (\ref{e37}) was obtained in a rather
conservative manner. The comparison of data from columns 2 and 5 of
Table~3 shows that even if the confidence interval were to be widened
to achieve 99\% confidence probability, the quantities
$\langle P_1^{\rm theor}-P^{\rm expt}\rangle$
would still remain outside this interval
within some separation region.
To make this argument  quantitative we
calculate the half-width of a new confidence interval from the equality
\begin{equation}
\frac{\Delta_{0.99}^{\!{\rm tot}}\left(P_1^{\rm theor}-
P^{\rm expt}\right)}{\Delta_{0.95}^{\!{\rm tot}}\left(P_1^{\rm theor}-
P^{\rm expt}\right)}=
\frac{t_{[1+0.99]/2}(2)}{t_{[1+0.95]/2}(2)}\approx 2.31.
\label{e40}
\end{equation}
\noindent
(Being conservative, we preserve only two degrees of freedom, i.e.
the minimum value from the analysis of Section 4.1).
Using the data from column 2 in Table~3 for
$\Delta_{0.95}^{\!{\rm tot}}\equiv \Delta^{\!{\rm tot}}$, one obtains
$\Delta_{0.99}^{\!{\rm tot}}\left(P_1^{\rm theor}
-P^{\rm expt}\right)=3.67\,$mPa,
1.46\,mPa, and 1.13mPa at separations $z=300\,$nm, 400\,nm, and
500\,nm, respectively. Comparing these results with the values of
$\langle P_1^{\rm theor}-P^{\rm expt}\rangle$ from column 5
in the same table, one concludes that the quantities
$\langle P_1^{\rm theor}-P^{\rm expt}\rangle$
still fall outside the new confidence interval in the separation
region $300\,\mbox{nm}\leq z\leq 500\,$nm.
We conclude that the first theoretical approach to the thermal
Casimir force  is excluded
experimentally with 99\% confidence.

Finally, we compare  the second
theoretical approach \cite{27} with the experimental data.
In Fig.~14b, we plot the differences of the Casimir pressures
$\left[P_2^{\rm theor}-P^{\rm expt}\right]$
versus separation $z$  for all 14
sets of measurements, as in Figs.~13 and 14a.
The solid lines show the
half-width of the confidence interval
$\Delta^{\!{\rm tot}}\left(P_2^{\rm theor}-P^{\rm expt}\right)
\approx
\Delta^{\!{\rm tot}}\left(P_4^{\rm theor}-P^{\rm expt}\right)$
at 95\%
confidence (there are no points above the $z$-axis in the
interval $160\,\mbox{nm}\leq z\leq 420\,$nm in the
approach of  \cite{27}).
As is seen from Fig.~14b, within the separation region
$160\,\mbox{nm}\leq z\leq 350\,$nm almost all points
$\left[P_2^{\rm theor}-P^{\rm expt}\right]$
fall outside the confidence interval.
It follows that
the second theoretical approach
is excluded experimentally at 95\% confidence.

This conclusion is confirmed by  considering the mean values
$\langle P_2^{\rm theor}-P^{\rm expt}\rangle$
obtained from the results from all
14 sets of measurements.
In column 6 of Table~3 the mean differences
$\langle P_2^{\rm theor}-P^{\rm expt}\rangle$
are presented at different separations
from 170\,nm to 700\,nm using the data from all 14
sets of measurements. The comparison of these data
with the half-width of the confidence interval in column 2
shows that at separations $170\,\mbox{nm}\leq z\leq 350\,$nm
the mean differences of the theoretical and experimental Casimir
pressures fall outside the confidence interval.
This confirms the conclusion that
the second theoretical approach is excluded by
the results of the Casimir pressure measurements using
 a micromachined oscillator at 95\% confidence.

\section{Stronger constraints on long-range interactions}

As mentioned in the Introduction,  measurements of the Casimir force
between metals have been successfully used to obtain
stronger constraints on hypothetical long-range interactions. Such
interactions have long been predicted in elementary particle physics
(see literature cited in  \cite{11}). They can arise from the exchange
of light elementary particles (scalar axions, graviphotons, dilatons,
and moduli among others \cite{56,57}) predicted by many extensions of
the Standard Model, and as a consequence of extra-dimensional theories
with low compactification scales \cite{58}. In most cases the potential
energy between two point masses $m_1$ and $m_2$ separated by
a distance $r$ is given by
the usual Newtonian potential with a Yukawa correction
\begin{equation}
V(r)=-\frac{Gm_1m_2}{r}\left( 1+\alpha
e^{-r/\lambda}\right),
\label{e42}
\end{equation}
\noindent
where $G$ is the gravitational constant, $\alpha$ is a dimensionless
constant characterizing the strength of the Yukawa force, and
$\lambda$ is its interaction range.

There have been many proposals to constrain
$\alpha$ and $\lambda$ over sub-$\mu$m
 interaction ranges from  van der Waals and
Casimir force measurements between macroscopic bodies (see  references
on the subject in  \cite{59}). During the last few years
the previously known constraints on ($\alpha,\lambda$) were strengthened
by up to 4500 times from  modern Casimir force measurements between
metals \cite{13,14,15,16,17}. In  \cite{11} an additional strengthening
by up to a factor of 11  was achieved
from the first precise Casimir pressure
measurement by means of a micromachined oscillator \cite{10,11}.

The results presented above from improved measurements of the Casimir
pressure compared with different theoretical approaches
 allow us to improve
the constraints found in  \cite{11}.
Previous limits have been strengthened by up to a factor of 2,
and their application range widened, while increasing their reliability.
This has been made possible by the use of the above metrological analysis
which confirmed the agreement between experiment and two
traditional theoretical approaches
within a wide separation range. (In previous experiments constraints were
usually obtained at some fixed separation distance.)

As in  \cite{11}, the Newtonian gravitational force between the plate
and the sphere of a micromachined oscillator,
and the equivalent gravitational
pressure between two plates are negligible. To calculate the equivalent
Yukawa pressure between two plates, one needs the detailed structure of
the sphere and plate materials. In the experiment under consideration
an Al${}_2$O${}_3$ sphere of
density $\rho_{AlO}=4.1\times 10^3\,$kg/m${}^3$ was coated with a
layer of Ti of thickness $\Delta_{Ti}=10\,$nm with
$\rho_{Ti}=4.51\times 10^3\,$kg/m${}^3$, and  a layer of Au of
thickness $\Delta_{Au}^{\! s}=200\,$nm with
$\rho_{Au}=19.28\times 10^3\,$kg/m${}^3$. The Si plate of density
$\rho_{Si}=2.33\times 10^3\,$kg/m${}^3$ was first coated with
a layer of Pt of thickness $\Delta_{Pt}=10\,$nm, with
$\rho_{Pt}=21.47\times 10^3\,$kg/m${}^3$,
and then with a layer of Au of
thickness $\Delta_{Au}^{\! p}=150\,$nm.
Considering that the conditions
$z,\,\lambda\ll R,\,L$ are satisfied (where $L=3.5\,\mu$m
is the thickness of the plate), the equivalent hypothetical pressure
between two parallel plates with the above layer structure
is \cite{11}
\begin{eqnarray}
&&
P^{\rm hyp}(z)=-2\pi G\alpha\lambda^2e^{-z/\lambda}
\nonumber \\
&&
\phantom{aa}
\times
\left[\rho_{Au}-\left(\rho_{Au}-\rho_{Ti}\right)e^{-\Delta_{Au}^{\! s}/\lambda}
-\left(\rho_{Ti}-\rho_{AlO}\right)e^{-(\Delta_{Au}^{\! s}+
\Delta_{Ti})/\lambda}
\right]
\label{e43}\\
&&
\phantom{aa}
\times
\left[\rho_{Au}-\left(\rho_{Au}-\rho_{Pt}\right)e^{-\Delta_{Au}^{\!p}/\lambda}
-\left(\rho_{Pt}-\rho_{Si}\right)e^{-(\Delta_{Au}^{\! p}+
\Delta_{Pt})/\lambda}
\right].
\nonumber
\end{eqnarray}
\noindent
Note that surface roughness, which was substantially reduced in this
experiment, cannot significantly affect the magnitude of a hypothetical
force with an interaction range of about 100\,nm.

We can now obtain constraints on the hypothetical Yukawa pressure
from the  agreement between our measurements of the Casimir pressure
and theory at 95\% confidence. According to our results, no deviations
between traditional
theories and experiment were observed, i.e., the hypothetical
pressure should be less than or equal to the half-width of the confidence
interval
\begin{equation}
| P^{\rm hyp}(z)|\leq\Delta^{\!{\rm tot}}
\left[P^{\rm theor}(z)-P^{\rm expt}(z)\right].
\label{e44}
\end{equation}
\noindent
Note that Eq.~(\ref{e44}) is metrologically reliable, since it does not
use the root-mean-square deviation as a measure of agreement between
theory and experiment to obtain constraints on hypothetical interactions.
It is also worthwhile to note that the Yukawa-type hypothetical
interaction depends on the separation distance quite differently
than the thermal Casimir pressures of the first and second
approaches, which were excluded
experimentally in Section 7.2. Because of this the mimicry of one
phenomenon by another is extremely unlikely.

The numerical analysis of Eqs.~(\ref{e43}) and (\ref{e44}) leads to the
conclusion that the strongest constraints are obtained within the
interaction region $40\,\mbox{nm}\leq\lambda\leq 370\,$nm from the
measurement data at separations from $z=210\,$nm (where the half-width of
the confidence interval is 6.89\,mPa) to $z=450\,$nm (where the half-width
is equal to 0.53\,mPa). In fact, with the increase of $\lambda$,
the strongest constraints on $\alpha$ are obtained from
the measurement data at larger separations. The resulting constraints
on $\alpha$ are plotted in Fig.~15 for different values of $\lambda$
(line 1a). In the same figure constraints from earlier experiments
are also shown. These were obtained from previous experiments with a
micromachined oscillator (line 1b), from old measurements of the
Casimir force between dielectrics \cite{12} (line 2), from Casimir force
measurements by means of a torsion pendulum \cite{2,13} (line 3), and by
the use of an atomic force microscope \cite{6,16} (line 4).
In all cases the region in the ($\alpha,\lambda$) plane above the line
is excluded, and below the line is allowed by the experimental results.
Note that the constraints from our experiment, line 1a,
are found at 95\% confidence.
For the previous constraints, given by lines 1b, 2, 3, and 4, the
confidence levels were not determined.

As is seen from Fig.~15, the present experiment leads to the strongest
constraints in the
interaction range $40\,\mbox{nm}\leq\lambda\leq 370\,$nm
which is wider than in the experiment of  \cite{11}.
Comparing the constraints from lines 2 and 4, the largest
improvement (by a factor
of 20) is achieved at $\lambda\approx 150\,$nm. If we compare the
present results with those of  \cite{11}, the largest improvement,
by a factor of 2.2, is achieved at $\lambda\approx 316\,$nm.
It should be noted that for the first time the new constraints, given by
line 1a, completely fill in the gap between the modern constraints
obtained by the atomic force microscope (line 4) and those obtained
using a torsion pendulum (line 3). Furthermore,
as was noted above, the constraints
of line 1a are determined at the 95\% confidence level, which makes them
the most reliable constraints obtained to date from  measurements
of the Casimir force.

\section{Conclusions and discussion}

In this paper we have presented the results of an improved experiment
on the dynamical determination of the Casimir pressure between two
plane parallel Au coated plates using a micromachined oscillator.
(We recall that the pressure between two plane plates was inferred from
the Casimir force between a plate and a sphere using the proximity
force theorem.)
Many improvements in the previously performed experiment of  \cite{11}
were made yielding a large dividend in precision,
accuracy and reliability of the
 results. The chief advantages of the new experiment lie in a
great improvement of the surface roughness, and a decrease
by a factor of 1.7 in
the error in determining of the absolute separations. The metrological
analysis of all experimental errors was performed at 95\% confidence.
In doing so, the data were analyzed for the presence of outlying
measurements. This permitted us to find the total experimental error
as a function of separation for subsequent comparison of
experiment and theory.

To utilize all the advantages of the improved measurements, the Casimir
pressures between two Au plates were calculated in the framework of
four theoretical approaches proposed in literature
based on the Lifshitz formula at nonzero
temperature, with the reflection coefficients expressed in terms of
both the surface impedance and the dielectric function. The complete
optical data for Au were utilized to take into account both the finite
conductivity and the thermal corrections to the Casimir force. In
doing so, many relevant factors and properties of Au films were
computed or estimated, such as the variation of the tabulated
optical  data due to the grain structure of a metal, and the influence of
spatial nonlocality and patch potentials. The surface roughness was
carefully taken into account including the role of
nonmultiplicative and diffraction-type effects. The metrological
analysis of all theoretical errors was performed at 95\%
confidence (including the error resulting from use of the
proximity force theorem, finiteness of the plate area, and errors
in the determination of the surface separations). As a result, the
total theoretical error (independent of the experimental one) was
combined using the metrological criteria.

To compare different theoretical approaches and experiment,
we replaced the root-mean-square
deviation previously used in literature, by a rigorous procedure for
the composition of the experimental and theoretical errors at 95\%
confidence. As a result, a confidence interval was found
for the random variables
$\left[P_i^{\rm theor}(z)-P^{\rm expt}(z)\right]$,
where $i=1,\,2,\,3$ and 4 denotes the different theoretical
approaches described in Sections 1,\,5. It was demonstrated that with
the impedance approach and the plasma model the mean values
$\langle P_{3,4}^{\rm theor}(z)-P^{\rm expt}(z)\rangle$,
computed at all separations,
fall inside the confidence interval. As a consequence, these approaches
are consistent with
experiment at 95\% confidence. In terms of the relative
uncertainty, there is  agreement between each of them and experiment
at the level (1.4--1.9)\% when the separation  changes in
the interval $170\,\mbox{nm}\leq z\leq 420\,$nm. This is the best
agreement achieved to date at 95\% confidence compared to any other
experiment measuring the Casimir force.
Despite the fact that the thermal Casimir force computed within the
impedance approach or the plasma model
 was found to be consistent with experiment, the
thermal correction by itself has not yet been directly measured.
According to these approaches, at $T=300\,$K
this correction is small, in qualitative agreement with case of
ideal metals, and can be readily measured in the near future by
means of proposed experiments \cite{60,61}.

The experimental results were used as
a test for the alternative approaches to the thermal Casimir force
which predict large thermal corrections at short separations.
The above metrological comparison between
experiment and theory was repeated
with the other values $P_1^{\rm theor}(z)$,
computed in accordance with the first approach \cite{25,54,55}.
 It was found that at all separations from $z=170\,$nm
to $z=700\,$nm the mean values
$\langle P_1^{\rm theor}(z)-P^{\rm expt}(z)\rangle$
fall outside the confidence
interval. Moreover, at separations $300\,\mbox{nm}\leq z\leq 500\,$nm
the mean Casimir pressures computed in the first approach
fall outside the even wider 99\% confidence interval.
This enabled us to conclude that the first theoretical approach to the
thermal Casimir force is
excluded by our measurements at 99\% confidence.

The same metrological comparison was carried out with respect to
the second approach \cite{27}. The
Casimir pressures $P_2^{\rm theor}(z)$ were computed
and compared with experiment.
It was shown that the mean values
$\langle P_2^{\rm theor}(z)-P^{\rm expt}(z)\rangle$
fall outside the confidence
interval for
separations $170\,\mbox{nm}\leq z\leq 350\,$nm. Thus, the second
theoretical approach is also
excluded experimentally at 95\% confidence.

Finally, the strong agreement between experiment and traditional
theory of the Casimir force was used to obtain more stringent
constraints on the
hypothetical Yukawa-type interactions predicted in high energy
physics. Using the results of two experiments (this one and of
 \cite{11}) the
constraints on $\alpha$ as a function of $\lambda$ were strengthened
by up to a factor of
20   within a wide interaction range.

\section*{Acknowledgments}
The authors would like to thank H.~B.~Chan for technical assistance
in sample preparation.
G.L.K. and V.M.M. are grateful to T.~N.~Siraya for several helpful
discussions on the metrological procedures of data analysis.
They are also grateful to Purdue University for kind hospitality
and financial support.
R.S.D. acknowledges financial support from the Petroleum Research
Foundation through ACS-PRF No. 37542--G. The work of E.F. is
supported in part by the U.S. Department of Energy under
Contract No. DE--AC02--76ER071428.


\newpage
\begin{table}[h]
\caption{Casimir pressures and different contributions to them.
See text for further discussion.}
\begin{tabular}{llllllll}
\multicolumn{1}{c}{Separation}&\multicolumn{3}{c}{Impedance approach} &
\multicolumn{3}{c}{Plasma model} &\multicolumn{1}{c}{\ } \\
$z$\,(nm) & $P_4^{(l=0)}$\,(Pa) &$P_4^{(l\geq 1)}$\,(Pa) &$P_4$\,(Pa) &
$P_3^{(l=0)}$\,(Pa) &$P_3^{(l\geq 1)}$\,(Pa) &$P_3$\,(Pa) &
$P_L^{(l\geq 1)}$\,(Pa)  \\
\hline
160 & --0.0715 & --1.0726 & --1.1441 & --0.0719 & --1.0430 & --1.1149 &
--1.07715 \\
200 & --0.03834& --0.47096 & --0.5093 & --0.03847 & --0.46335 & --0.5018 &
--0.4724\\
250 & --0.02046& --0.20421 &--0.2247 & --0.02050 & --0.20260 & --0.2231 &
--0.2046 \\
300 & --0.01220 & --0.10214 &--0.1143 & --0.01222 &--0.10182 & --0.1140 &
--0.1023  \\
350 & --0.007858& --0.056424 & --0.06428 &--0.007866 &--0.056425 &
--0.06429 &
--0.05648 \\
400 &  --0.005358&--0.033547 &--0.03890 &--0.005362 & --0.033616&
--0.03898 &
--0.03357 \\
450 & --0.003817&--0.021105 &--0.02492&--0.003820& --0.021178& --0.02500&
--0.02111 \\
500 & --0.002816 & --0.013887 &--0.01670& --0.002817& --0.013948 &
--0.01676 & --0.01389 \\
600 & --0.001659 & --0.006665&--0.008324&--0.001660&--0.006703&--0.008363&
--0.006667 \\
700 & --0.001059 & --0.003546&--0.004605&--0.001059&--0.003569&
--0.004628&
--0.003547 \\
\end{tabular}
\end{table}
\vspace*{1cm}
\begin{table}[h]
\caption{Casimir pressures and zero-frequency contributions to them
in the alternative approaches to the thermal Casimir force.}
\begin{tabular}{lllll}
\multicolumn{1}{c}{Separation}&\multicolumn{2}{c}{Approach 1 \cite{25,54,55}} &
\multicolumn{2}{c}{Approach 2 \cite{27}}  \\
$z$\,(nm) & $P_1^{(l=0)}$\,(Pa) &$P_1$\,(Pa) &
$P_2^{(l=0)}$\,(Pa)  &$P_2$\,(Pa) \\
\hline
160 & --0.04836 & --1.1255 & --0.0967  & --1.1739 \\
200 & --0.0247 & --0.4971 & --0.0495 & --0.5219\\
250 & --0.0127 & --0.2173 & --0.0254 & --0.2300 \\
300 & --0.00734 & --0.1096 & --0.0147 & --0.1170 \\
350 & --0.00462 & --0.06110 &--0.00924 &--0.06572 \\
400 &  --0.00310 &--0.03667 &--0.00619 & --0.03976 \\
450 & --0.00217 & --0.02329 &--0.00434 & --0.02546 \\
500 & --0.00158 & --0.01547 & --0.00317 & --0.01706 \\
600 & --0.000917& --0.007584&--0.001835 &--0.008502 \\
700 & --0.000577& --0.004124 &--0.001155&--0.004702 \\
\end{tabular}
\end{table}
\begin{table}[h]
\caption{Mean values of differences between the theoretical and
experimental Casimir pressures. See the text
for the definition of $\Delta\bar{P}_i$.}
\begin{tabular}{cccccc}
$z$& $\Delta^{\!{\rm tot}}\left[P^{\rm theor}(z)-
P^{\rm expt}(z)\right]$ &
$\Delta\bar{P}_4$&$\Delta\bar{P}_3$&
$\Delta\bar{P}_1$ &
$\Delta\bar{P}_2$ \\
(nm) & (mPa) & (mPa) & (mPa) & (mPa) & (mPa)\\
\hline
170 & 17.2 & 2.01 & 13.0& 18.8 & --21.8 \\
180 & 13.4 & --0.74 & 7.54 & 14.4 & --19.8 \\
200 & 8.59 & --1.21 & 5.3& 11.0 & --13.9  \\
250 & 3.34 & --0.31 & 1.3& 7.09 & --5.66\\
300 & 1.59 & 0.34 & 0.6& 5.07 & --2.28  \\
350 & 0.89 & 0.38 & 0.39& 3.58 & --1.05 \\
400 & 0.63 & 0.28 & 0.20& 2.59 & --0.62 \\
500 & 0.49 & 0.11 & 0.05& 1.37 & --0.22 \\
600 & 0.46 & 0.08 & 0.04& 0.82 &--0.09 \\
700 & 0.46 & 0.02 & --0.01& 0.51 & --0.07 \\
\end{tabular}
\end{table}
\begin{figure}[h]
\caption{\label{fig1}Schematic diagram showing the experimental
setup. The different distances are defined in the text.}
\end{figure}

\begin{figure}[h]
\caption{\label{AFM}AFM images of the Au films deposited on the
MTO and on the sphere. Topographies on the different panels are
indicated. $10\times 10\,\mu\mbox{m}^2$ images of the film on the
MTO before the experiment (a) and after it (b) are provided.
Figures (c) and (d) provide similar information for the sphere,
except that a $5\times 5\,\mu\mbox{m}^2$ area is shown.}
\end{figure}

\begin{figure}[h]
\caption{\label{f3}
Typical cross sections of the atomic force microscope
 images of the Au coatings on the plate (a) and on the sphere (b).
$h$ denotes the height of the surface above the reference level
defined in the text,  and is plotted against the lateral position
$x$. }
\end{figure}

\begin{figure}[h]
\caption{\label{f4} Topographic heights $h_i$ on the sample as a
function of the total sample area with heights $h < h_{i+1}$ for
the film deposited on the MTO (a) and the  film deposited
on the sphere (b).
For both films the differences observed when analyzing
different regions in the sample are negligible.}
\end{figure}

\begin{figure}[h]
\caption{\label{deltaV}Dependence of the angular deviation
$\theta$ as a function of  the applied voltage to the sphere. Data
obtained at five different separations $z$ between the metallic
layers are shown. Data have been displaced vertically for the sake
of clarity.}
\end{figure}

\begin{figure}[h]
\caption{\label{Fel}Electrostatic force $F_{el}$ as a function of
separation $z$ for $V_{Au} -V_0$ = 0.22\,V, 0.25\,V 0.35\,V.
For simplicity, the $z$-axis has already been corrected for $D$
and $\theta$. The lines are fits to the data using Eq.~(\ref{eqe1}).}
\end{figure}

\begin{figure}[h]
\caption{\label{freq}Change in resonant frequency
$\Delta f_r=(\omega_r-\tilde{\omega}_r)/(2\pi)$
as a function of amplitude of
motion of the MTO, obtained at a separation $z = 300$\,nm. The
non-linear behavior becomes dominant at large amplitudes, increasing the
uncertainty in the frequency. Frequency shifts are measured with respect
to the one obtained using the thermodynamic noise as a driving force,
see text. Inset: Resonance curve obtained under the conditions indicated by
the arrow.
}
\end{figure}

\begin{figure}[h]
\caption{\label{data} Absolute value of the parallel plate Casimir
pressure $P^{\rm expt}$ as a function of separation $z$.  The
separation has been corrected taking into account $D$ and $\theta$.
}
\end{figure}

\begin{figure}[h]
\caption{\label{f9} The random $\Delta^{\!{\rm rand}}P^{\rm expt}$
(long-dashed lines), systematic
$\Delta^{\!{\rm syst}}P^{\rm expt}$ (short-dashed
lines) and total $\Delta^{\!{\rm tot}}P^{\rm expt}$ (solid line)
absolute experimental errors of the Casimir pressure
measurements versus separation within the intervals
$160\,\mbox{nm}\leq z\leq 750\,$nm (a) and
$160\,\mbox{nm}\leq z\leq 750\,$nm (b).}
\end{figure}

\begin{figure}[h]
\caption{\label{f11} The total relative error of the Casimir pressure
measurements versus separation.}
\end{figure}

\begin{figure}[h]
\caption{\label{f12}  The relative theoretical errors of the
Casimir pressure computed in different approaches
due to use of the proximity force theorem
(short-dashed line), variation of the tabulated optical  data
(long-dashed line), due to the experimental uncertainties of
absolute separations (dotted line), and total theoretical relative
error (solid line) versus separation.}
\end{figure}

\begin{figure}[h]
\caption{\label{f13} The 95\% confidence intervals
(solid lines) and differences
$\left[P_4^{\rm theor}-P^{\rm expt}\right]$ (dots) versus
separation within the intervals $160\,\mbox{nm}\leq z\leq 250\,$nm (a)
and $250\,\mbox{nm}\leq z\leq 750\,$nm (b) for one set of
measurements computed in the impedance approach.}
\end{figure}

\begin{figure}[h]
\caption{\label{f14} The 95\% confidence intervals
(solid lines) and differences
$\left[P_4^{\rm theor}-P^{\rm expt}\right]$ (dots) versus
separation within the intervals $160\,\mbox{nm}\leq z\leq 250\,$nm (a)
and $250\,\mbox{nm}\leq z\leq 750\,$nm (b) for all fourteen sets of
measurements computed in the impedance approach.}
\end{figure}

\begin{figure}
\caption{\label{f15} The 95\% confidence intervals
(solid lines) and differences
$\left[P_{1,2}^{\rm theor}-P^{\rm expt}\right]$ (dots) versus
separation for theoretical approaches 1 \cite{25,54,55} (a)
and 2 \cite{27} (b) incorporating different models of alternative
thermal corrections. All fourteen sets of
measurements as in Fig.~13 were used.}
\end{figure}
\newpage
\begin{figure}[h]
\caption{\label{f16} Constraints on the strength of the Yukawa interaction
$\alpha$ versus interaction range $\lambda$.
Line 1a is obtained in this paper, line 1b is from  \cite{11},
line 2 follows from old Casimir force measurements
between dielectrics \cite{12}. Lines 3 and 4 are obtained from
Casimir force measurements by using
a torsion pendulum \cite{2,13}, and by means of an atomic force
microscope \cite{6,16}, respectively. The region of ($\alpha,\lambda$)
plane above each line is excluded and below each line
is allowed.}
\end{figure}

\end{document}